\documentclass[%
 reprint,
nofootinbib,
 amsmath,amssymb,
 aps,
pra,
floatfix,
]{revtex4-2}

\usepackage{graphicx}
\usepackage{dcolumn}
\usepackage{bm}
\usepackage{hyperref}
\usepackage[mathlines]{lineno}
\usepackage{braket}
\usepackage{overpic}
\usepackage{tikz}

\newcommand{\del}{{\rm d}}

\newcommand{\trdev}[2]{\partial_{\langle #1}\partial_{#2 \rangle}}

\usepackage{xcolor}

\begin{document}

\preprint{APS/123-QED}

\title{Fully-relativistic evolution of vacuum tensor inhomogeneities during inflation}

\author{Ericka Florio}
\email{eaf49@cam.ac.uk}
\author{E. Paul S. Shellard}%
 \email{eps1@cam.ac.uk}
\affiliation{%
 Centre for Theoretical Cosmology, Department of Applied Mathematics and Theoretical Physics,
University of Cambridge, Wilberforce Road, Cambridge CB3 0WA, United Kingdom
}%

\date{\today}

\begin{abstract}
We present a complete method for the initialisation and extraction of first-order inflationary tensor perturbations for fully relativistic simulations which incorporate gravitational backreaction. 
We outline a correspondence between the Cosmological Perturbation Theory (CPT) framework and the numerical relativity BSSN variables in the appropriate limit. 
We describe a generation method for stochastic tensoral initial conditions, inspired by the standard scalar initial condition used from inflation and implemented in lattice cosmology. 
We discuss the implementation of this procedure in the GRChombo/GRTeclyn code, and demonstrate the detailed quantitative correspondence between the linearised and fully-nonlinear solutions in the perturbative limit, through the evolution of the background and the tensor power spectrum.  We also validate the methodology by showing that energy and momentum constraints are introduced and preserved to second-order or better. We provide some preliminary indicative results probing tensoral non-Gaussianity using the skewness and kurtosis. The computational pipeline presented here will be used to study the emergence of a primordial tensor bispectra and cross-spectra that incorporate the effect of nonlinear gravitational couplings with the metric, which has potential applications for the analysis of next-generation CMB surveys.
\end{abstract}

\maketitle

\section{Introduction}
\label{sec:intro}
Single-field slow-roll inflation, when added to the $\Lambda$CDM model, explains many properties of the CMB signal, such as its phase coherence and Gaussianity. 
However, the most standard single-field models has been heavily constrained by the combined BiCEP/Planck experiments (see Refs.~\cite{tsujikawaPlanckConstraintsSinglefield2013,martinBestInflationaryModels2014,bicep/keckcollaborationImprovedConstraintsPrimordial2021}, among others).
This result has renewed interest in the investigation of non-standard models of inflation, particularly into their distinguishing observable signals.

A primordial stochastic gravitational wave background (PGWB) is expected to be produced by most inflationary models, whether minimally via slow-roll inflation (Ref.~\cite{starobinskiiSpectrumRelictGravitational1979}), or through nonlinear mechanisms such as preheating (Ref.~\cite{eastherGravitationalWaveProduction2007}).
These primordial tensoral perturbations are model-dependent, and therefore carry information about the evolution of inflaton that is complementary to scalar perturbations.
A stochastic gravitational wave background (SGWB) has already been detected by the NANOGrav collaboration (Ref.~\cite{agazieNANOGrav15Yr2023}). 
Although its origins remain unclear, this signal could have been generated by early Universe processes (see for example Refs.~\cite{drewAxionStringSource2024,tadaMultifieldStochasticDynamics2024,choudhuryPrimordialNonGaussianitySaviour2024,launayStochasticInflationGeneral2024a}).

The PGWB, if detected, would provide key hints about the nature of the inflationary epoch.
For instance, the amplitude of the primordial tensor power spectrum is directly related the Hubble factor $H$ during the inflationary period, and would serve as a measure of the energy scale of inflation (Ref.~\cite{baumannCosmology2022}).
To date, the most robust constraint on the amplitude of PGWB produced by standard inflation was measured by the Planck 2018 data release and the BiCEP experiment. 
Combined, these data sets were able to constrain the error in the tensor-to-scalar ratio, $\delta r$, to less than $0.03$ at a $95\%$ confidence level (Ref.~\cite{paolettiPlanckBICEPKeck2022}). 
The BiCEP collaboration has now released data which gives an uncertainty of $\sigma(r) < 0.009$ (Ref.~\cite{bicep/keckcollaborationLatestConstraintsInflationary2022}), and the Simons Observatory, which saw its first light this year, is projected to measure $\sigma(r)$ on the order of $10^{-4}$ (Ref.~\cite{abazajianCMBS4ForecastingConstraints2022a}).
Additionally, the LiteBIRD experiment, set to launch in 2028, will be able to push current measurements of the scalar to tensor ratio below $0.002$ at $95$\% confidence level (Ref.~\cite{litebirdcollaborationProbingCosmicInflation2023}).

In addition to the SGWB, higher-order statistics in scalar (and potentially in tensor) fields are also promising observables which can drastically constrain the space of inflationary models, if they are detected.
Here we focus on the primordial bispectrum, which is expected to be generated for all but the simplest models of inflation.
The bispectrum can theoretically be measured in the scalar field alone, using the three-point scalar correlator $\langle\zeta\zeta\zeta\rangle$; in the scalar and tensor field together, via for example $\langle\zeta\zeta\gamma\rangle$; or in the tensor field alone, via $\langle\gamma\gamma\gamma\rangle$.
The primordial bispectrum is particularly useful in distinguishing multi-field models \cite{bernardeauNonGaussianityMultifieldInflation2002}, which can provide a high-quality power spectrum fit while introducing a unique bispectrum signature.

Both the PGWB and the primordial bispectrum have the potential to constrain the current parameter space of inflationary models, but only if we have developed robust predictions for their spectral shapes in advance.
Historically, these spectra have computed using the linearised equations of motion as the sources for the inflaton field and the metric (for example, Ref.~\cite{chenLargeNonGaussianitiesIntermediate2010}), or using lattice simulations which resolve the nonlinear dynamics of the matter sector, but assume the gravitational sector evolves according to linear theory (Ref.~\cite{dufauxTheoryNumericsGravitational2007a,dufauxGravityWavesTachyonic2009,priceStochasticBackgroundsGravitational2008,garcia-bellidoStochasticBackgroundGravitational2007a,khlebnikovRelicGravitationalWaves1997}).
However, when perturbations to the metric couple at beyond leading order, as can happen in multi-field inflation (see Ref.~\cite{iacconiMultifieldInflationLarge2023} for example), it is reasonable to expect that beyond-linear-order dynamics in both the matter sector and the metric may become important in the final shape of these spectra. 

Many numerical algorithms have been developed recently which can fully capture all aspects of these nonlinear signals. 
This new discipline of lattice cosmology has produced a variety of codes (Refs.~\cite{felderLATTICEEASYProgramLattice2008,giblinGravitationalRadiationPreheating2010,huangArtLatticeGravity2011,cloughGRChomboNumericalRelativity2015a,figueroaCosmoLatticeModernCode2023}), which have been employed most often to study (p)re-heating scenarios (see for example Refs.~\cite{giblinGravitationalRadiationPreheating2010,vandevisTimeScalesNonlinear2020,adsheadGravitationalWavesKinetic2024a,figueroaGeometricReheatingUniverse2024}) and phase transitions (for example, Ref.~\cite{dahlDecayAcousticTurbulence2022}).
Some lattice codes focus on parametric resonance in the matter sector, generated during multi-particle inflation and preheating (Refs.~\cite{figueroaGeometricReheatingUniverse2024,figueroaNonlinearDynamicsAxion2024,nguyenNonlinearDynamicsPreheating2019,bastero-gilNonlinearPreheatingScalar2008}). 
In this case, nonlinear interactions between the various matter fields are all well-resolved, however the back-reaction onto the metric is assumed to be negligible.
In a few cases, a fully-nonlinear evolution of both the metric \textit{and} the matter fields has been employed, for instance to study primordial black hole formation during preheating scenarios (Refs.~\cite{giblinjrPreheatingFullGeneral2019,joanaCosmicInhomogeneitiesEarly2022,adsheadGaugePreheatingFull2024}).

However, the use of fully-relativistic lattice simulations to make predictions from the inflationary epoch \textit{itself} has been limited. 
This is in part due to the great success of linear theory in describing dynamics during inflation. 
However, multi-field models are known to produce potentially large non-Gaussian signals where there is a strong bend in the field-space trajectory (Ref.~\cite{chenLargeNonGaussianitiesIntermediate2010,bernardeauNonGaussianityMultifieldInflation2002,raveendranNumericalEvaluationTensor2017}), a scenario which numerical tools are well-suited to study.
Additionally, second-order coupling between scalar and tensor perturbations has been shown to produce unique spectral shapes in the PGWB (Ref.~\cite{bariGravitationalWavesInduced2024}).

In this work, we use the Numerical Relativity (NR) code GRChombo to fully evaluate the dynamics of the metric and the inflaton during inflation.
GRChombo has previously been used to show that inflating patches are generically generated from large scalar field fluctuations (Ref.~\cite{cloughRobustnessInflationLarge2018,cloughDifficultyGeneratingGravitational2018}) and various potential shapes (Ref.~\cite{aurrekoetxeaEffectsPotentialShape2020,joanaGravitationalDynamicsHiggs2022,joanaCosmicInhomogeneitiesEarly2022}). 
We will focus here on a detailed, quantitative study of the inflationary epoch itself, both incorporating and improving similar tools developed by others in the lattice cosmology community (see for example Ref.~\cite{figueroaArtSimulatingEarly2021}) to connect our numerical results with perturbative methods arising from the quantum treatment of the inflationary problem (Refs.~\cite{khlebnikovClassicalDecayInflaton1996,maldacenaNongaussianFeaturesPrimordial2003a}).
We have developed a distinct example of GRChombo specifically designed to study the deep inflationary regime, which we are actively porting to the GRTeclyn code.

This work is organised as follows. 
In Sec.~\ref{sec:cosmo-theory}, we review the relevant theory of gravitational waves propagation on Minkowski and FLRW backgrounds, as well as the quantisation of these perturbations.
In Sec.~\ref{sec:rel-defs}, present a gauge-agnostic dictionary which can be used to translate first-order cosmological perturbations to and from the fully-relativistic BSSN variables in the linear regime. 
In Sec.~\ref{sec:numerical-methods}, we describe our particular numerical implementation in GRChombo,\footnote{We expect our implementation in GRTeclyn will follow this implementation closely.} including our choice of gauge and the initial conditions for the background spacetime and perturbations. 
In Sec.~\ref{sec:cosmo-evo}, we present validation results for the evolution of the background in the super-horizon and horizon-crossing regimes, and of the evolution of the tensor power spectrum. 
We demonstrate the constraint-satisfying properties of our initial data, and provide an improvement to the standard stochastic initialisation used in the lattice cosmology literature.
We show convergence test results, and give a first look into how this program could capture the dynamic emergence of higher-order correlators.
In Sec.~\ref{sec:summary}, we summarise our results and outline the immediate next steps of our research program.

Throughout this work we use the mostly-plus signature of the metric, $\eta_{\mu\nu} = \mbox{diag}(-1,+1,+1,+1)$, and natural units\footnote{For more information on how $m_{pl}$ is used to set the scale of spacetime in GRChombo, see Appendix A in Ref.~\cite{cloughScalarFieldsNumerical2018}.} $c=\hbar=1,$ giving $G=m_{pl}^{-2}.$ When translating the background equations of motion into the GRChombo/GRTeclyn program, we will define ``program units'' which absorb this factor of $m_{pl}$ into our physical quantities, making them unitless.\footnote{For instance, in QFT units the background inflaton field $\bar{\phi}$ has units of mass, and so the corresponding ``program'' field given by $\bar{\phi}_{pr} \equiv \bar{\phi}/m_{pl}.$}

\section{Primordial gravitational waves}
\label{sec:cosmo-theory}

\subsection{Gravitational waves in Minkowski spacetime}
\label{subsec:gw-in-mink}
Gravitational waves represent two free, propagating degrees of freedom in the metric. They are traditionally formulated in terms of a perturbation on the Minkowski metric,
\begin{eqnarray*}
    g_{\mu\nu} = \eta_{\mu\nu} + h_{\mu\nu},
\end{eqnarray*}
where the evolution equations of the perturbations $h_{\mu\nu}$ follow from Einstein's equations.
In order to simplify the evolution equations, it is common to consider the trace-reversed metric perturbation (Ref.~\cite{carrollLectureNotesGeneral1997}), given by
\begin{eqnarray*}
    \tilde{h}_{\mu\nu} = h_{\mu\nu} - \frac{1}{2} \eta_{\mu\nu} h.
\end{eqnarray*}

One must also choose a gauge, which fixes the degrees of freedom associated with the choice of coordinates. The typical gauge choice in Minkowski space involves first taking the partial gauge-fixing known as the de-Donder gauge (Ref.~\cite{carrollLectureNotesGeneral1997}), where
\begin{eqnarray}
    \label{eqn:de-donder}
    \partial^{\mu} h_{\mu\nu} - \frac{1}{2} \partial_{\nu} h = 0
\end{eqnarray}
Decomposing onto plane-wave solutions, one can then show that the remaining degrees of freedom can be eliminated by prescribing certain components of the plane-wave solution. 
The typical choice here is the transverse-traceless (TT) condition (Ref.~\cite{carrollLectureNotesGeneral1997}), which sets
\begin{equation}
    h_{0\mu} = h^{\mu}_{\mu} = 0.
\end{equation}
In this gauge, the metric perturbation is fully spatial and the full metric can be simplified to
\begin{equation}
    \label{eqn:gw-mink-metric}
    \del s^2 = -\del t^2 + (\delta_{ij} + h_{ij}) \del x^i \del x^j.
\end{equation}
The spatial metric perturbation $h_{ij}$ is traceless and transverse in spatial derivatives, meaning $h^i_i = \partial^i h_{ij} = 0$.
Under these gauge conditions, it can then be shown that in vacuum Einstein's equations reduce to the massless wave equation for $\tilde{h}_{\mu\nu}$ (Ref.~\cite{carrollLectureNotesGeneral1997}):
\begin{equation*}
    \Box \tilde{h}_{\mu\nu} = 0.
\end{equation*}

In the TT gauge, the metric perturbation contains only two degrees of freedom, which we choose to decompose into plus and cross polarisations. 
Consider a gravitational wave propagating along the \textbf{k} direction, and consider two vectors \textbf{m} and \textbf{n} which form an orthonormal basis together with \textbf{k}. 
One can then construct two polarisation basis tensors, given by
\begin{eqnarray}
    \nonumber \epsilon^{+}_{ij} &=& \textit{m}_i \textit{m}_j - \textit{n}_i \textit{n}_j\\
    \label{eqn:pol-tensors}
    \epsilon^{\times}_{ij} &=& \textit{m}_i \textit{n}_j + \textit{n}_i \textit{m}_j,
\end{eqnarray}
which obey the orthonormality condition
\begin{equation*}
    \sum_{i,j} \epsilon^s_{ij} \epsilon^{s'}_{ij} = 2\delta^{ss'}
\end{equation*}
where $s,s'\in \{+,\times\}$, and where the sum does \textit{not} assume symmetric indexing on $i$ and $j$ (Ref.~\cite{isiParametrizingGravitationalwavePolarizations2022}). The gravitational wave perturbation can then be constructed from these basis tensors as 
\begin{equation}
    \label{eqn:plus-cross-reconstruction}
    h_{ij} = h_+ \epsilon^+_{ij} + h_{\times} \epsilon^{\times}_{ij} = \sum_s h_s \epsilon^s_{ij}
\end{equation}
where $h_+$ and $h_{\times}$ represent the two true degrees of freedom in the metric (Ref.~\cite{baumannCosmology2022}), which we call the \textit{mode functions} of the metric. 
These mode functions can be recovered from the spatial metric using the orthonormality condition,
\begin{equation*}
    h_s = \frac{1}{2} \sum_{i,j} h_{ij} \epsilon^s_{ij}.
\end{equation*}


\subsection{Gravitational waves in cosmological spacetimes}
\label{subsec:cpt}

\subsubsection{The inflationary background}

Homogeneous, isotropic and flat spacetimes are described generally by the FLRW metric,
\begin{eqnarray*}
    \del s^2 = -\del t^2 + a^2(t)\delta_{ij}\del x^i \del x^j.
\end{eqnarray*}
Applying this metric to Einstein's equations, assuming spatial flatness and linearising the time-time component gives the Friedman equation for the metric, 
\begin{eqnarray}
    \label{eqn:fried}
    H^2 &=& \frac{8\pi}{3 m_{pl}^2}\rho.
    \end{eqnarray}
To generate a generic inflationary spacetime, we assume the matter energy density $\rho$ arises from a scalar inflaton field $\phi$, which evolves according to the simplified Klein-Gordon equation
    \begin{eqnarray}
    \label{eqn:kg}
    \ddot{\phi}  - 3H\dot{\phi} - \frac{\del V}{\del \phi} = 0.
\end{eqnarray}
The energy density of the inflaton field is given by the 00 component of the energy-momentum tensor,
\begin{eqnarray}
\label{eqn:rho-def}
    \rho = \frac{1}{2}\dot{\phi}^2 + V(\phi).
\end{eqnarray}

Deep in the inflationary regime we define an early epoch of \textit{slow-roll}, where the dimensionless slow-roll parameters (SRPs)\footnote{Note that we do \textit{not} assume that the slow-roll assumption holds in our definition of these parameters, and instead use their most ``general'' form which also applies in nonlinear regimes and at late times during inflation.}
\begin{eqnarray}
    \label{eqn:srps}
    \epsilon \equiv -\frac{\dot{H}}{H^2}, \
    \delta \equiv -\frac{\ddot{\phi}}{H\dot{\phi}}
\end{eqnarray}
are small compared to 1. 
In this regime, $\ddot{\phi} \ll H\dot{\phi}$ and the Klein-Gordon equation simplifies to
\begin{eqnarray}
\label{eqn:kg-slow-roll}
    3H\dot{\phi} + \frac{\del V}{\del \phi} = 0,
\end{eqnarray}
which, together with Eqn.~\eqref{eqn:fried} and \eqref{eqn:rho-def}, forms a complete set of differential equations which can be solved for the background solution.

\subsubsection{The gauge-agnostic CPT formalism}

Various tools developed for the flat-space formulation of gravitational waves will aid us in developing an understanding of tensoral perturbations of cosmological spacetimes. 
Cosmological GWs propagate on a quasi-de-Sitter background, whereas astrophysical GWs propagate on nearly Minkowski spacetime. 
As de-Sitter spacetime has an inherent preferred time direction, whereas Minkowski does not, the formulation of the TT gauge for PGWs be carefully considered.\footnote{For a review of how NR gauges have been adapted to study cosmological spacetimes, see Ref.~\cite{aurrekoetxeaCosmologyUsingNumerical2024}.}
We describe PGWs using the Cosmological Perturbation Theory (CPT) decomposition of the metric (see for example Ref.~\cite{baumannCosmology2022}).
The CPT metric is based on a scalar-vector-tensor decomposition of perturbations evolving on an FLRW background, and formulates these perturbations in a gauge-agnostic manner.
 We follow the notation used in Ref.~\cite{baumannCosmology2022}, and begin by perturbing each component of the metric:
\begin{eqnarray}
    \nonumber \del s^2 = -&\bar{N}^2 &(1 + 2A) \del t^2 + 2a \bar{N} B_{i} \del t \del x^i \\
    \label{eqn:CPT-simple} &+& a^2(\delta_{ij} + 2E_{ij}) \del x^i \del x^j
\end{eqnarray}
where $\bar{N}$ chooses the time slicing of the background.\footnote{Commonly one chooses $\bar{N}=1$ (cosmic time) or $\bar{N}=a$ (conformal time)}  
This formulation captures all possible first-order metric perturbations.

The perturbations $B_i$ and $E_{ij}$ contain multiple types of degrees of freedom -- for instance, $B_i$ contains an inherent scalar (magnitude) and vector (direction) perturbation.
We decompose $B_i$ and $E_{ij}$ into their inherent scalar, vector and tensor degrees of freedom (Ref.~\cite{baumannCosmology2022}):
\begin{eqnarray*}
    B_i &=& \partial_i B + \hat{B}_i\\
    E_{ij} &=& C\delta_{ij} + \trdev{i}{j} E + \partial_{(i} \hat{E}_{j)} + \hat{E}_{ij}
\end{eqnarray*}
where $|\hat{B}_i|=|\hat{E}_i|=1$, where $\trdev{i}{j}$ represents the trace-free derivative of $E$,
\begin{equation*}
    \trdev{i}{j} E = \left(\partial_i \partial_j - \frac{1}{3} \nabla^2 \delta_{ij}\right)E,
\end{equation*}
and where $\hat{E}_{ij} \equiv 1/2\ h_{ij},$ as given by Eqn.~\eqref{eqn:gw-mink-metric}.
We can see that the CPT metric truly contains four scalar perturbations ($A,B,C,E$), two vector perturbations ($\hat{B}_i, \hat{E}_i$), and one tensor perturbation, $\hat{E}_{ij}$.

It is well known that in inflationary spacetimes, linear vector perturbations evolve along decaying solutions (Refs.~\cite{golovnevCosmologicalPerturbationsVector2009,baumannCosmology2022}). 
Thus we will neglect vector perturbations of the metric by assuming they are vanishing on our initial slice, and use the scalar-tensor form of the CPT metric given by
\begin{eqnarray}
    \nonumber \del s^2 = -\bar{N}^2 (1 + 2A) \del t^2 + 2a \bar{N} (\partial_i B)\del t \del x^i \\
    \label{eqn:CPT-ST} + a^2 [(1 + 2C)\delta_{ij} + 2 \trdev{i}{j} E + 2 \hat{E}_{ij}]\del x^i \del x^j.
\end{eqnarray}

The purpose of this present work is to construct a complete methodology for initialising and extracting tensor perturbations from cosmological NR simulations. 
As we mention in Sec.~\ref{sec:summary}, we plan to extend this work to cover both scalar and tensoral perturbations in the near future (see for example Ref.~\cite{launayStochasticInflationGeneral2024a}), and will present a complete dictionary between scalar and tensor CPT perturbations and BSSN variables here.
However, for all of the numerical results presented in this work, we do not initialise scalar perturbations, and thus the initial metric is given simply by
\begin{eqnarray}
    \label{eqn:CPT-T}
    \del s^2 &=&-\bar{N}^2 \del t^2 + a^2(\delta_{ij} + h_{ij})\del x^i \del x^j.
\end{eqnarray}
Note that this corresponds to the geodesic gauge choice in NR, which is equivalent to synchronous gauge in cosmological perturbation theory (Ref.~\cite{aurrekoetxeaCosmologyUsingNumerical2024}).


\subsection{Vacuum tensor perturbations}
\label{subsec:ics-formulation}

The transverse-traceless part of the spatial metric can be thought of as a massless spin-two field -- a graviton. 
If we assume that the graviton field is in its ground state and that the field was in a Minkwoskian vacuum state in the far past, then its mode functions will evolve according to the Mukhanov-Sasaki equation (Ref.~\cite{baumannCosmology2022}). 
In this section I will briefly describe the results of the Mukhanov-Sasaki equation, which we will use to generate a stochastic realisation of the graviton field on the lattice.

\subsubsection{Fourier convention}

The Mukhanov-Sasaki formulation is written in Fourier space. 
We write the transform from Fourier space to configuration space as
\begin{eqnarray*}
    f(\textbf{x}) = \int d^3k f(\textbf{k})e^{-i \textbf{k}\cdot \textbf{x}}
\end{eqnarray*}
and the transform from configuration space to Fourier space as
\begin{eqnarray*}
    f(\textbf{k}) = \frac{1}{(2\pi)^3} \int d^3x f(\textbf{x})e^{i \textbf{k}\cdot \textbf{x}}.
\end{eqnarray*}
In GRChombo, we use the discrete definition for each component of $\textbf{k}$ given by:
\begin{eqnarray*}
    k_j = \frac{2\pi j}{L}
\end{eqnarray*}
where $L$ is the length of the box in program units, and $j$ is the unitless index running from $[0, N-1]$, where $N$ gives the total number of points on one edge of the lattice.

\subsubsection{The Mukhanov-Sasaki solution}

The Einstein-Hilbert action in natural units is given by 
\begin{eqnarray*}
    S = \frac{m_{pl}^2}{16 \pi} \int \del^4 x \ R.
\end{eqnarray*}
Applying the tensoral CPT line element, and writing the tensoral perturbation as
\begin{eqnarray}
    \label{eqn:plus-cross-ds-reconstruction}
    f_{ij} = \sum_s f_s \epsilon^s_{ij}
\end{eqnarray}
where $f_s \equiv a h_s$ is the FLRW mode function analogous to the gravitational mode functions in Minkowski space, we can derive the first-order tensoral contribution to the action as
(Refs.~\cite{Mukhanov:1990me,Kodama:1984ziu}) 
\begin{eqnarray*}
    S^T = \frac{m_{pl}^2}{16\pi} \sum_{s\in[+,\times]} \int \del \tau \del^3 x \left[(f_s')^2 - (\nabla f_s)^2 + \frac{a''}{a}f_s^2\right].
\end{eqnarray*}
By varying this action we arrive at the Mukhanov-Sasaki equation, which in Fourier space reads
\begin{equation}
    \label{eqn:ms-confml-time}
    f''_{s} + \left(k^2 - \frac{a''}{a}\right)f_{s} = 0.
\end{equation}
We note that we can transform this equation such that it solves for $h_s$ rather than $f_s$, and from conformal time to cosmic time, which gives
\begin{equation}
    \label{eqn:h-cosmic-kg}
    \ddot{h}_{s} + 3H\dot{h}_{s} + \left(\frac{k}{a}\right)^2 h_{s} = 0.
\end{equation}

In the sub-horizon limit, where $a''/a  \ll k^2 $, Eqn.~\eqref{eqn:ms-confml-time} becomes a wave equation for $f_s$. 
Thus in this regime, we can promote the tensoral degrees of freedom $f_s$ to quantum operators
\begin{equation*}
    \hat{f}_s({\bm{k}}) = f_s(k)\hat{a}_{\bm{k}} + f^*_s(k)\hat{a}^{\dagger}_{-\bm{k}},
\end{equation*}
where the quantum mode functions $f_s$ satisfy the classical wave equation (Ref.~\cite{khlebnikovClassicalDecayInflaton1996}). The vacuum state for a quantum operator $\hat{f}_s$ satisfying this equation is the Bunch-Davies vacuum
\begin{equation*}
    f_s(k, \tau) = \frac{e^{-ik\tau}}{m_{pl}\sqrt{2k}},
\end{equation*}
where we have used $\omega^2 = k^2$, corresponding to a massless Bunch-Davies state, and where we have chosen the positive-frequency solution. 
This can be used as an initial condition for Eqn.~\eqref{eqn:ms-confml-time}.
If we assume small slow-roll parameters, we can approximate $a''/a \sim 2/\tau^2,$ giving the solution to Eqn.~\eqref{eqn:ms-confml-time} in the slow-roll regime as
\begin{equation}
    \label{eqn:MS-soln-field}
    f_s(k, \tau) = \frac{e^{-ik\tau}}{m_{pl}\sqrt{2k}}\left(1-\frac{i}{k\tau}\right).
\end{equation}
Taking the conformal time derivative of this mode function gives the appropriate solution for the velocity field,
\begin{equation}
    \label{eqn:MS-soln-velocity}
    f'_s(k,\tau) = \frac{e^{-ik\tau}}{m_{pl}\sqrt{2k}}\left(\frac{i}{k\tau^2} - ik -\frac{1}{\tau}\right).
\end{equation}
We will use Eqns.~\eqref{eqn:MS-soln-field} and \eqref{eqn:MS-soln-velocity} to build stochastic initial conditions for the metric.

\section{Gravitational waves in Numerical Relativity}
\label{sec:rel-defs}
We wish to use the perturbative evolution of tensor perturbations to inform the development of our numerical simulations, where the interaction between various metric degrees of freedom can contribute at any order. 
First, we formulate the perturbative CPT variables in terms of the non-perturbative BSSN variables, in the regime where the two sets can be identified.
This will allow us to phrase the initial conditions of our simulation in terms of a semi-classical realisation of the CPT variables, and to create a basic extraction algorithm for these variables.

\subsection{The BSSN formalism}
\label{subsec:bssn}

The BSSN formalism (Refs.~\cite{baumgarteNumericalIntegrationEinsteins1998,shibataEvolutionThreedimensionalGravitational1995a}) is a particular decomposition of the ADM (Ref.~\cite{arnowittDynamicsGeneralRelativity2004}) form of Einstein's equations, which casts Einstein's equations into a Cauchy problem for the spatial metric on a set of hypersurfaces $\Sigma_t$. The BSSN equations are found by plugging the ADM form of the metric,
\begin{equation}
    \label{eqn:adm-metric}
    \del s^2 = -\alpha^2 \del t^2 + \gamma_{ij}(\del x^i + \beta^i \del t)(\del x^j + \beta^j \del t)
\end{equation}
or, equivalently,
\begin{equation}
    \label{eqn:adm-cpt-comp}
    \del s^2 = (-\alpha^2 + \beta_i \beta^i)\del t^2 + 2\beta_i \del x^i \del t + \gamma_{ij} \del x^i \del x^j.
\end{equation}
into Einstein's equations. Note that $\gamma_{ij}$, the spatial metric, is typically chosen to represent all physical degrees of freedom, and $\alpha$ (the lapse) and $\beta^i$ (the shift) represent the gauge degrees of freedom. 

We also need the definition of the extrinsic curvature, which takes the form
\begin{equation}
    \label{eqn:extrinic-curve-def}
    K_{ij} = -\frac{1}{2\alpha}(\partial_t \gamma_{ij} - D_{(i} \beta_{j)} )
\end{equation}
on the hypersurface $\Sigma_t$, where $D_i$ is the covariant 4-derivative projected onto the hypersurface by $\gamma_{ij}$. This is the first ADM equation. Applying Eqn.~\eqref{eqn:adm-metric} and \eqref{eqn:extrinic-curve-def} to Einstein's equations gives the second ADM equation,
    \begin{eqnarray}
    \nonumber \partial_t K_{ij} &=& \alpha(R_{ij} - 2K_{ik}K^{k}_j + KK_{ij})\\\nonumber &-& D_i D_j \alpha + \beta^k\partial_k K_{ij} + K_{k(i}\partial_{j)}\beta^k \\ 
    \label{eqn:adm-evo-eqns:extrinsic-curve} &-& \frac{8\pi\alpha}{m_{pl}^2} (S_{ij} - \tfrac{1}{2}\gamma_{ij}(S-\rho)),
\end{eqnarray}
where $R_{ij}$ are the components of the Ricci tensor on the spatial hypersurface, and $S_{ij}$, $S$ and $\rho$ are derived from the stress-energy tensor (Refs.~\cite{baumgarteNumericalRelativitySolving2010,radiaLessonsAdaptiveMesh2022}). 
These equations of motion are accompanied by a set of constraint equations, which must be satisfied on every hypersurface, and are given by
\begin{eqnarray}
    \label{eqn:constraints-ham}
    R + K^2 + K_{ij}K^{ij} &=& 16\pi G \rho\\
    \label{eqn:constraints-mom}
    D_j(K^{ij} - \gamma^{ij} K) &=& 8\pi G S^i.
\end{eqnarray}

The BSSN decomposition defines a separate ``scalar'' and ``tensor'' component for the $\gamma_{ij}$ and $K_{ij}$ variables, and reformulates Eqns.~\eqref{eqn:extrinic-curve-def} and \eqref{eqn:adm-evo-eqns:extrinsic-curve} in terms of these new components. 
This reformulation transforms the ADM equations into a well-posed system of PDEs (Ref.~\cite{baumgarteNumericalRelativitySolving2010}).
The spatial metric is decomposed into a conformal factor $\chi \equiv (\mbox{det}[\gamma_{ij}])^{-1/3}$ and a conformally-rescaled metric $\tilde{\gamma}_{ij}$ such that
\begin{equation}
    \label{eqn:confml-metric-def}
    \gamma_{ij} = \frac{1}{\chi}\tilde{\gamma}_{ij}
\end{equation}
and $\mbox{det}[\tilde{\gamma}_{ij}] = 1$ (Ref.~\cite{radiaLessonsAdaptiveMesh2022}). The extrinsic curvature is decomposed onto a trace $K \equiv K^i_i$ and traceless metric $\tilde{A}_{ij}$ such that
\begin{equation}
    K_{ij} = \chi\left(\tilde{A}_{ij} - \frac{1}{3}\tilde{\gamma}_{ij} K\right)
\end{equation}
and $\tilde{A}^i_i = 0$  (Ref.~\cite{radiaLessonsAdaptiveMesh2022}).
Applying this decomposition to Eqns.~\eqref{eqn:extrinic-curve-def} and \eqref{eqn:adm-evo-eqns:extrinsic-curve} yields a set of four coupled first-order PDEs, called the BSSN equations:
\begin{eqnarray*}
    \partial_t \chi &=& \beta^k\partial_k \chi + \frac{2}{3}\chi(\alpha K - \partial_k \beta^k)\\
    \partial_t \tilde{\gamma}_{ij} &=& \beta^k \partial_k \tilde{\gamma}_{ij} + \tilde{\gamma}_{k(i}\partial_{j)}\beta^k - 2\alpha \tilde{A}_{ij} - \frac{2}{3}\tilde{\gamma}_{ij}\partial_k\beta^k\\
    \partial_t K &=& \beta^k \partial_k K + \alpha(R + K^2) - \gamma^{kl}D_k D_l \alpha \\ & & \hspace{0.4\linewidth} + \frac{4\pi\alpha}{m_{pl}^2}(S - 3\rho)\\
    \partial_t \tilde{A}_{ij} &=& \beta^k \partial_k \tilde{A}_{ij} + \chi \left(-D_i D_j \alpha + \alpha(R_{ij} - \tfrac{8\pi}{m_{pl}^2} S_{ij})\right)^{TF}\\ &+&\tilde{A}_{ij}\left ( \alpha K - \frac{2}{3} \partial_k \beta^k \right) + 2 \tilde{A}_{k(i} \partial_{j)} \beta^k - 2 \alpha 
 \tilde{\gamma}^{kl} \tilde{A}_{ik} \tilde{A}_{jl}
\end{eqnarray*}
where $(-)^{TF}$ represents the trace-free part of the quantity in brackets, and where $R = \gamma^{ij}R_{ij}$  (Ref.~\cite{radiaLessonsAdaptiveMesh2022}).
The variables which must be specified on the initial slice of the simulation are: $\{\chi, \tilde{\gamma}_{ij}, K, \tilde{A}_{ij}, \alpha, \beta^i\}.$ 
 The lapse and the shift will be initialised according to our gauge choice, but the remainder will require physically-motivated initial data.

The constraint equations \eqref{eqn:constraints-ham} and \eqref{eqn:constraints-mom} can be re-written in terms of these BSSN variables as 
\begin{eqnarray}
    \label{eqn:constr-ham-bssn}
    {\cal H} &\equiv& R + \frac{2}{3}K^2 - \tilde{A}_{ij}\tilde{A}^{ij} - 16\pi G \rho = 0\\
    \nonumber{\cal M}_i &\equiv& \tilde{\gamma}^{kl}\left( \partial_k \tilde{A}_{il} - 2\tilde{\Gamma}^m_{l(i}\tilde{A}_{k)m} - 3 \tilde{A}_{ik} \frac{\partial_l \chi}{2\chi} \right)\\
    \label{eqn:constr-mom-bssn} & & \hspace{0.3\linewidth} - \frac{2}{3} \partial_i K - 8\pi G S_i = 0,
\end{eqnarray}
where I have introduced the BSSN constraint-tracking variables ${\cal H}, {\cal M}_i$ (Ref.~\cite{radiaLessonsAdaptiveMesh2022}).\footnote{Note that in this work, ${\cal H}$ does \textit{not} refer to the conformal-time Hubble parameter.} 

\subsection{The CPT-BSSN correspondence}
\label{subsec:cpt-bssn-dict}

We present here a complete correspondence between the BSSN and CPT variables in the linear regime. 
This correspondence has been noted previously for specific gauge choices, for example in Ref.~\cite{giblinjrPreheatingFullGeneral2019} for the Newtonian gauge.
However, we have developed this correspondence in a \textit{gauge-agnostic} manner, such that any particular gauge choice can be applied directly to the correspondence equations. 
We anticipate that this dictionary may facilitate comparisons between the results of lattice calculations which have been formulated in separate gauges.

This correspondence was made by comparing the perturbative metric, Eqn.~\eqref{eqn:CPT-ST}, to the fully-relativistic metric, Eqn.~\eqref{eqn:adm-cpt-comp}, and by calculating the extrinsic curvature to first order in perturbative variables. 
First, we can assign the shift term according to the $\del t \del x^i$ component of Eqn.~\eqref{eqn:CPT-ST},
\begin{eqnarray*}
    \beta^i = 2a\bar{N}\partial_i B.
\end{eqnarray*}
Then the $\beta^i\beta_i$ term in Eqn.~\eqref{eqn:adm-cpt-comp} is second order in perturbations, and can be ignored. We assign the lapse to be
\begin{equation*}
    \alpha = \bar{N}\sqrt{1+2A} \approx \bar{N}(1+A).
\end{equation*}
where the approximation denotes a Taylor expansion truncated at first order. The spatial metric is assigned simply as
\begin{equation}
    \label{eqn:gamma-cpt}
    \gamma_{ij} = a^2 [(1 + 2C)\delta_{ij} + 2 \trdev{i}{j} E + 2 \hat{E}_{ij}].
\end{equation}
The extrinsic curvature can now be calculated according to Eqn.~\eqref{eqn:extrinic-curve-def}, where we note that at zeroth order the Christoffel coefficients vanish, and as $\beta_i$ is strictly first-order in perturbations, we can replace $D_i \rightarrow \partial_i$. This gives
\begin{eqnarray}
    \nonumber K_{ij} &=& a^2\biggl[H\left( A\delta_{ij} - \frac{\gamma_{ij}}{a^2}\right) + \frac{1}{\bar{N}}\left(\partial_t \frac{\gamma_{ij}}{a^2}\right) - \frac{1}{a}\partial_i \partial_j B \biggr] \\
     \nonumber &=& a^2 \biggl[H\left((A - 2C - 1)\delta_{ij} - 2\partial_{<i}\partial_{j>}E - 2\hat{E}_{ij}\right)\\ 
     \label{eqn:kij-down}  \ & & \hspace{0.07\linewidth} + \frac{1}{\bar{N}}( \dot{C}\delta_{ij} + \partial_{<i}\partial_{j>}\dot{E} + \hat{\dot{E}}_{ij}) - \frac{1}{a}B_{,ij}\biggr].
\end{eqnarray}

We can now calculate the BSSN variables.  Using the fact that $\trdev{i}{j} E$ and $\hat{E}_{ij}$ are traceless, the conformal factor becomes
\begin{equation}
    \label{eqn:chi-corresp}
    \chi  = \frac{1}{a^2}(1+6C)^{-\frac{1}{3}} \approx \frac{1}{a^2}(1-2C).
\end{equation}
Then the conformally rescaled metric is 
\begin{equation}
    \label{eqn:gamma-corresp}
    \tilde{\gamma}_{ij} = \chi \gamma_{ij} \approx \delta_{ij}+ 2\partial_{<i}\partial_{j>}E + 2\hat{E}_{ij}.
\end{equation}
The trace of the extrinsic curvature is calculated by finding $\gamma^{ij}$ to first order, raising the index of $K_{ij}$, and then taking the trace. This gives
\begin{equation}
    \label{eqn:K-corresp}
     K \approx -3H(1 - A) -3\frac{\dot{C}}{\bar{N}} - \frac{1}{a}\nabla^2 B,
\end{equation}
where we have again used the fact that $\trdev{i}{j} E$ and $\hat{E}_{ij}$ are trace free. The trace-free extrinsic curvature follows simply:
\begin{eqnarray}
    \nonumber \tilde{A}_{ij} &=& \frac{1}{\chi}K_{ij} + \frac{1}{3}\tilde{\gamma}_{ij}K\\
    \label{eqn:Aij-corresp} &\approx&  \frac{1}{\bar{N}}\left[\partial_{<i}\partial_{j>}\left(\dot{E} + \frac{B}{a}\right)  -\hat{\dot{E}}_{ij}\right].
\end{eqnarray}

Equations \eqref{eqn:chi-corresp}, \eqref{eqn:gamma-corresp}, \eqref{eqn:K-corresp} and \eqref{eqn:Aij-corresp} represent the full translation between CPT and BSSN variables, in the case where scalar and tensor perturbations are included. 
We note also that this work is compatible with the work presented in Ref.~\cite{launayStochasticInflationGeneral2024a} for the case of stochastic evolution of scalar perturbations during inflation.
However, as we will be concerned primarily with tensoral perturbations in this work, we neglect scalar perturbations, giving the tensor-only correspondence:
\begin{eqnarray}
    \nonumber \chi &=& \frac{1}{a^2}\\
    \nonumber \tilde{\gamma}_{ij} &=& \delta_{ij} + 2\hat{E}_{ij} =\delta_{ij} + h_{ij} \\
   \nonumber K &=& -3H \\
   \label{eqn:tensor-only-corresps}
    \tilde{A}_{ij} &=& -\frac{1}{\bar{N}}\hat{\dot{E}}_{ij} = -\frac{1}{2\bar{N}}\dot{h}_{ij}.
\end{eqnarray}
Thus any method for prescribing the cosmological variables $\{a, H, h_{ij}, \dot{h}_{ij}\}$ is sufficient to initialise the BSSN variables in the linear regime.

If we apply the correspondence Eqn.~\eqref{eqn:tensor-only-corresps} to the Hamiltonian constraint, Eqn.~\eqref{eqn:constr-ham-bssn}, we find
\begin{eqnarray*}
    {\cal H} &=& R(h) + 6H^2 - \frac{1}{4}\dot{h}_{kl}\dot{h}^{kl} - 16\pi G \rho = {\cal O}(h^2),
\end{eqnarray*}
since $R$ and $\tilde{A}^{ij}\tilde{A}_{ij}$ only receive contributions at second order and above. Applying \eqref{eqn:tensor-only-corresps} to \eqref{eqn:constr-mom-bssn} gives
\begin{eqnarray*}
   \nonumber {\cal M}_i &=& -\frac{1}{2}\delta^{kl} \left(\partial_k \dot{h}_{li} - 2\tilde{\Gamma}^m_{l(i}\dot{h}_{k)m}\right) - 8\pi G S_i,
\end{eqnarray*}
and given that $\tilde{\Gamma}^i_{jk}$ is first-order in perturbations, that $S_i = 0$ and that $\dot{h}_{ij}$ is transverse, this gives ${\cal M}_i = {\cal O}(h^2)$. Thus as long as we detect constraint violations that are suppressed to second-order, we can be confident in our first-order results.

\section{Numerical methods and implementation}
\label{sec:numerical-methods}
In this section, we will provide details for how our initial data was constructed, and how the tensorial perturbations were extracted from GRChombo.

\subsection{Gauge choice}
\label{subsec:gauge}

The gauge choice in GRChombo is written in the general Bona-Masso family of slicing conditions, first formulated in Ref.~\cite{bonaNewFormalismNumerical1995}, and is represented by an initial condition and first-order equation of motion for the gauge variables $\alpha$ and $\beta^i$. We write the Bona-Masso condition as
\begin{eqnarray*}
    \label{eqn:alpha-eom}
    \partial_t \alpha &=& a_1 \beta^k \partial_k \alpha - a_2 \alpha^{a_3}K\\
    \label{eqn:beta-eom}
    \partial_t \beta^i &=& b_1 \beta^k \partial_k \beta^i + b_2 B^i \\
    \label{eqn:B-eom}
    \partial_t B^i &=& b_1(\beta^k \partial_k B^i - \beta^k \partial_k \tilde{\Gamma}^i) + \partial_t \tilde{\Gamma}^i - \eta B^i.
\end{eqnarray*}
where $\tilde{\Gamma}^i \equiv \tilde{\gamma}^{jk}\tilde{\Gamma}^i_{jk}$ are the contracted Christoffel coefficients on the rescaled spatial metric, and where $\eta$ is a new gauge parameter controlling the damping of $B^i$ (Ref.~\cite{radiaLessonsAdaptiveMesh2022}).

A particular gauge within the Bona-Masso family is selected by choosing $\{a_i, a_2, a_3, b_1, b_2, \eta\}.$ 
In this work, we must use geodesic gauge in either cosmic or conformal time, as this is required by our initial metric, Eqn.~\eqref{eqn:CPT-T}. 
We chose to use cosmic time, and set $\alpha = 1$ and $\beta^i = 0$ on the initial slice, and choose all Bona-Masso coefficients to be zero so that the lapse and the shift should not evolve off of these values.\footnote{As the Christoffel coefficients vanish to leading order for TT tensoral perturbations, $\partial_t B^i$ should remain zero as well. Any deviation from this path could indicate a departure from perfect TT gauge.}

We may wish to move to conformal time, where $\bar{N} = a(\tau)$ rather than 1. This can be accomplished in the Bona-Masso family by setting $\alpha = a^2(0)$ on the initial slice, and choosing $a_1 = 0, a_2 = 2/3$ and $a_3 = 1$. This ensures
\begin{eqnarray*}
    \partial_{\tau} \alpha = -\frac{2}{3}\alpha K \sim 2 a^2 H
\end{eqnarray*}
as expected, where $\alpha(\tau) = a^2(\tau)$.

\subsection{Initial data construction}
\label{subsec:ic-construction}

The background initial conditions are determined by the choice of inflaton model, and describe precisely which period of inflation we are evolving.
The perturbation initial conditions will be generated similarly to the standard initialisation scheme used in cosmological lattice simulations, which seeks to approximate the quantum nature of these perturbations at very large occupation number with a semi-classical stochastic realisation (see for example Refs.~\cite{giblinjrPreheatingFullGeneral2019,felderLATTICEEASYProgramLattice2008,figueroaArtSimulatingEarly2021}).

\subsubsection{Inflationary model choice}

The initial conditions of the background arises from the Friedman and Klein-Gordon equations.
We choose to start in a regime of slow-roll, where we can use the slow-roll Klein-Gordon equation, as outlined in Sec.~\ref{subsec:ic-construction}. 
One can combine Eqns.~\eqref{eqn:fried} and \eqref{eqn:kg-slow-roll} and, together with Eqn.~\eqref{eqn:rho-def} and a choice for the form of $V(\phi)$, evaluate these on the initial slice. 
Taking the large-field toy model $V = \frac{1}{2}m^2 \phi^2$, we find
\begin{eqnarray*}
    H_{0}^2 &=& \frac{8\pi}{3 m_{pl}^2}\left(\frac{1}{2} \dot{\phi}_0^2 + \frac{1}{2} m^2 \phi_0^2\right)\\
    3 H_0 \dot{\phi}_0 &=& -m^2 \phi_0.
\end{eqnarray*}

In order to ensure we are deep in the inflationary epoch, we choose  $\phi_0 = 4 \ m_{pl}$ with mass parameter $m = 10^{-4} \ m_{pl}$.
This choice is similar to the choice made in Ref.~\cite{cloughRobustnessInflationLarge2018}.
The initial Hubble parameter and scalar field velocity are found by solving the above equations, and are given by $H_0 \approx 8.1934\cdot 10^{-4} \ m_{pl}$ and $\dot{\phi}_0 \approx -1.6273\cdot 10^{-5}\ m_{pl}^2$. 
This set of parameters gives an initial SRP of $\epsilon_0 \approx 2\cdot 10^{-3}$.
Note that we feed these initial conditions into GRChombo with 15 decimal places of accuracy, in order to reduce any error present in the incorrect assignment of these values to the level of floating point accuracy.

\subsubsection{Gaussian random field generation}

Inflation is most often considered to be driven by a quantum field, the inflaton. 
However, parts of inflation may be made amenable to the techniques of classical field theory, where a ``quantum-to-classical'' transition may be justifiable. 
This transition is usually associated with modes crossing outside the horizon (see, e.g., Refs.~\cite{polarskiSemiclassicalityDecoherenceCosmological1996} or \cite{baumannCosmology2022}). 
For perturbations much smaller than the horizon size, it appears to be necessary to treat them as quantum, ensuring continuity with the Bunch-Davies vacuum state in the far past.
On the other hand, perturbations after horizon-crossing are frozen into the background spacetime on super-horizon scales (at least for slow-roll single-field inflation) and then follow entirely classical equations of motion.  However, as shown explicitly in Ref.~\cite{Launay:2024trh}, the classical approximation has wider application because there is an important distinction between the quantum initial conditions and the ensuing evolution of this vacuum state, which has both quantum and classical contributions.  Predicted observables like the power spectrum from single field inflation are reproduced using classical evolution on the initial BD vacuum state (e.g.\ \cite{Mukhanov:1990me}), while higher-order correlators like the bispectrum may also be dominated by evolutionary contributions that are classical rather than quantum, though this is a model-dependent statement~\cite{Launay:2024trh}.  

Another previously-studied quantum-to-classical transition is the end of inflation, where $\tau \rightarrow 0$ as reheating takes hold.
As argued in Ref.~\cite{khlebnikovClassicalDecayInflaton1996}, in this limit the mode function of the inflaton field will become a stochastic realisation of the underlying quantum system, and will obey fully-classical equations of motion. 
Refs.~\cite{giblinjrPreheatingFullGeneral2019,eastherStochasticGravitationalWave2006,eastherGravitationalWaveProduction2007,felderLATTICEEASYProgramLattice2008} use this fact to construct initial data for lattice simulations of the preheating period.
We will use a similar method to construct stochastic initial data for tensoral perturbations in this near-horizon regime.

Consider the quantum state whose mode functions obey the Mukhanov-Sasaki solution, Eqn.~\eqref{eqn:ms-confml-time}. 
The two-point correlator of this state is given by
\begin{equation*}
    \braket{ \hat{f}_s(\textbf{k}) \hat{f}_s(\textbf{k}')} \equiv \bra{0}\hat{f}_s(\textbf{k}) \hat{f}_s(\textbf{k}')\ket{0}.
\end{equation*}
and inserting the decomposition onto raising/lowering operators gives the relation to the power spectrum,
\begin{equation*}
  \braket{ \hat{f}_s(\textbf{k}) \hat{f}_s(\textbf{k}')} = (2\pi)^2 P_h(k) \delta (\textbf{k}-\textbf{k}').
\end{equation*}

In the case of reheating, oscillatory modes are most important to the dynamics of interest, as the interaction between these modes produces the resonance that characterises the thermalisation of the post-inflationary Universe (Refs.~\cite{giblinjrPreheatingFullGeneral2019,eastherGravitationalWaveProduction2007,aurrekoetxeaOscillonFormationInflationary2023,deskinsGaugeFieldPreheating2013,giblinGravitationalRadiationPreheating2010,battefeldSuppressionParametricResonance2009}). 
Thus in these cases, all modes are sub-horizon on the initial slice, in which case the power spectrum is well-approximated by the Bunch-Davies spectrum, $P_{BD} = 1/2\omega_k$.
Here, we are interested in modes that are very near the horizon, and that cross the horizon within the dynamic range of our simulation.
Therefore we will base our power spectrum on the full solution to the Mukhanov-Sasaki equation, represented by the square of the mode function, Eqn.~\eqref{eqn:MS-soln-field}:
\begin{eqnarray}
    \label{eqn:MS-power-spectrum}
    \nonumber P_h(k) &=& |h_k(\tau)|^2 = \frac{1}{a^2m_{pl}^2}|f_k(\tau)|^2\\
    \label{eqn:field-spec} &=& \frac{1}{a^2 m_{pl}^2} \left(\frac{1}{2k} + \frac{(aH)^2}{2k^3}\right).
\end{eqnarray}
Note here that we have used the slow-roll identity $\tau = aH$, as we know $H_0$ on the initial slice.
This power spectrum transitions smoothly between the expected power spectrum for sub-horizon modes, given by a $1/k$ scaling, to the spectrum for super-horizon modes, given by a $1/k^3$ scaling (Ref.~\cite{baumannCosmology2022}).

We will represent the tensor initial conditions as a Gaussian random field (GRF) which follows the power spectrum given in Eqn.~\eqref{eqn:field-spec}. 
GRFs can be constructed by decomposing the field at each point in Fourier space into a magnitude and phase,
\begin{eqnarray}
    \label{eqn:grf-mode-fn}
    h_s(k) = M_h e^{i\theta_h}.
\end{eqnarray}
Here, $M_h$ is drawn from a Rayleigh distribution\footnote{Note that, as C++ has no in-built function to draw from a Poisson distribution, we use an equivalent Poisson draw given by 
\begin{eqnarray*}
    R(P)=\sqrt{-2\ln(U)P(k)}.
\end{eqnarray*}
where $U \sim {\cal U}(0,1)$ is a random variable drawn from the standard uniform distribution.} with size parameter $\sigma_h = \sqrt{P_h(k)}$, and $\theta_{h}$ represents the phase of the quantum state.
Typically, $\theta_h$ has been simply represented by a random variable drawn from the uniform distribution ${\cal U}(0,2\pi)$. However, we have found that the inherent phase present in the mode function \eqref{eqn:MS-soln-field} is important when seeking to accurately recover the correct spectral evolution predicted by the Mukhanov-Sasaki equation \ref{eqn:ms-confml-time} (see Sec.~\ref{subsec:single-mode} for more details); this is essentially ensuring that the quantum state has the appropriate initial conditions for the classical evolution.  

The configuration-space polarisation fields are found via the inverse Fourier transform:
\begin{equation*}
    h_s(\textbf{x}) = \int d^3 k \ h_s(\textbf{k})e^{-i \textbf{k}\cdot \textbf{x}}\,,
\end{equation*}
with the reality of the initial conditions enforced with appropriate conditions on the complex conjugate. 
The Fourier-space tensor object can be constructed from the mode functions using the polarisation basis tensors,\footnote{See Apdx. \ref{apdx:plus-cross-construction} for further details on how these polarisation basis tensors are constructed in Fourier space.}
\begin{eqnarray*}
    h_{ij}(\textbf{k}) = \sum_s h_s(k)\epsilon^s_{ij}(\textbf{k}).
\end{eqnarray*}
The appropriate initial conditions for the rescaled spatial metric are then
\begin{eqnarray}
\label{eqn:hij-config}
    h_{ij}(\textbf{x}) = \int d^3k \ h_{ij}(\textbf{k}) e^{-i \textbf{k}\cdot \textbf{x}}
\end{eqnarray}
Note that $h_{ij}(\textbf{x})$ is not found by combining $h_s(\textbf{x})$ and a set of $\epsilon_{ij}$ constructed in configuration space. 
This is because the Fourier transform does not distribute over the multiplication of the mode functions and the basis tensors. 
In the same vein, the configuration-space mode functions are not be found by using the orthonormality condition on $h_{ij}(\textbf{x})$. 
The Fourier-space tensor object $h_{ij}(\textbf{k})$ is found first, in both cases.

We have developed a new initial condition class of GRChombo which generates this random tensor perturbation on the lattice.\footnote{We expect that this class will be made public in an upcoming release of GRChombo/GRTeclyn, however if you would like access to this functionality before such time, please contact the authors.}
This program also verifies that the tensor field is trace-free to a specified level of precision.
The polarisation fields $h_s$ are then extracted from GRChombo's output files, using a similar program to the initial condition class but which is currently external to GRChombo (but which shortly will be combined).  
An initial polarisation field generated by this class is shown in panel (a) of Fig.~\ref{fig:box-plots}.

To fully initialise GRChombo we need an initial profile for $\dot{h}_{ij}$. 
In order to derive the power spectrum of $\dot{h}$, we must translate our expression for the comoving tensor perturbation $f_s$
 into an expression for $h_s$, and we must change our time coordinate from conformal to cosmic time. Performing these operations, we find 
 \begin{eqnarray*}
     \dot{h}(t) &= \frac{1}{a^2}\left(f'(\tau) - \frac{a'}{a}f(\tau)\right)\\
     &= -\frac{i}{a^2}\sqrt{\frac{k}{2}}e^{-ik\tau}
 \end{eqnarray*}
 meaning that the power spectrum for the velocity field has the same scaling with $k$ as in Minkowski space:
\begin{eqnarray}
    \label{eqn:MS-PS-velocity}
    \nonumber P_{\dot{h}}(k) &=& |f'_k(\tau)|^2\\
    &=& \frac{1}{a^4 m_{pl}^2} \frac{k}{2}.
\end{eqnarray}
Note that this power spectrum also recovers the super-horizon limit, as where $k$ is very small, the power spectrum approaches zero.
The random velocity field is generated by taking the exact same random amplitude and phase draw as used in the position field, and applying this power spectrum to that draw.

\subsubsection{Window function}

Random initial data can have large spatial derivatives on small length scales. 
However,  our finite grid resolution limits the precision with which we can resolve the derivative between two nearby points. 
Initial data which is too ``noisy'' on small scales can introduce numerical instabilities early, and the derivative stencil will artificially inflate power in high modes as the simulation progresses.
A common method used to avoid this issue involves introducing a window function on the initial data (See Refs.~\cite{aurrekoetxeaOscillonFormationInflationary2023,giblinjrPreheatingFullGeneral2019}). 
This suppresses the power leakage into high modes by damping their power on the first slice.

We chose to use a tanh-like window function,
\begin{equation*}
    W_{k_*, \Delta_k}(k) = \frac{1}{2}[1 - \tanh(\Delta_k(k-k_*))],
\end{equation*}
where $k_*$ is the cut-off mode and $\Delta_k$ is the window width.
We measure $k_*$ in terms of the unit length of the isotropic power spectrum, $2\pi/L$, and measure $\Delta_k$ in terms of $L$.

\subsection{Extraction of transverse-traceless perturbations}
\label{subsec:extraction}

As long as we remain in the perturbative regime, we should be able to extract the tensor perturbations at successive slices using the simple relation
\begin{equation}
\label{eqn:extr-simple}
    h_{ij}(t) = \tilde{\gamma}_{ij}(t) - \delta_{ij}.
\end{equation}
We employ this scheme for the results shown here, and ensure that the perturbations remain at least trace-free with the same tests as were run with the initial condition generation class.
We note that this method avoids the complication laid out in Ref.~\cite{figueroaTransversetracelessProjectionLattice2011}, which details how the transverse nature of a gravitational wave signal evolved on the lattice depends on the spatial derivative stencil used in the evolution scheme.\footnote{We are, however, able to ensure our initial conditions are transverse on the first slice, before any spatial derivative stencil has altered them.}

However, if the metric were to develop a significant trace or transverse part, as will be the case when we include initial scalar perturbations, we would need a more robust extraction mechanism for tensor modes, which can adequately separate each component of the metric and which takes into account the derivative stencil used. 
We can construct an improved extraction method by using the decomposition, outlined clearly in Ref.~\cite{otaCovariantTransversetracelessProjection2022}, which explicitly removes the trace and the transverse component of a generic metric. 
Let $g_{ij}$ be a general perturbation of the full metric $\tilde{\gamma}_{ij}$, which may or may not be transverse-traceless.\footnote{In fact, on an FLRW background $g_{ij}$ would correspond to the tensor perturbation $E_{ij}$ presented in Eqn.~\eqref{eqn:CPT-simple}.}
We can write $g_{ij}$ as 
\begin{equation}
    g_{ij} = \frac{1}{3} \delta_{ij} g + 2 D^{(i} V^{j)} + h_{ij}
\end{equation}
where $g \equiv g^i_i$ and 
\begin{equation*}
    V^i \equiv \gamma^{il} \partial^m \left(g_{ml} - \frac{1}{3} \delta_{ml} g \right),
\end{equation*}
representing the transverse violation of the trace-free perturbation (Ref.~\cite{otaCovariantTransversetracelessProjection2022}).
Then the true transverse-traceless signal can be extracted most generally from the rescaled spatial metric $\tilde{\gamma}_{ij}$ by calculating 
\begin{equation}
\label{eqn:ext-full}
    h_{ij} = \tilde{\gamma}_{ij} - \delta_{ij} - \frac{1}{3} \delta_{ij} g - 2 D^{(i} V^{j)},
\end{equation}
where $D_i$ is approximated using the same stencil which was used in the evolution scheme. This reduces to the simple extraction method in the case where $g = V^i = 0$.
We expect to show further results using this improved method of extraction in an upcoming publication.

\section{Cosmological evolution and validation}
\label{sec:cosmo-evo}
\begin{figure}
    {\centering
\includegraphics[width=1.0\linewidth]{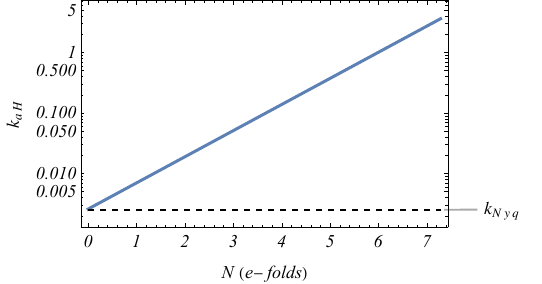}
    }
    \caption{The evolution of the Hubble mode $k_{aH}$ (blue) as a function of e-folds, in the super-horizon case. The black dashed line shows the smallest resolvable mode, $k_{Nyq}$, for a grid size $N=128$.}
    \label{fig:sup-horizon-evo}
\end{figure}

Here, we provide validation of our GRChombo example in the super-horizon horizon-crossing regime. 
We define these two cases according to the size of the mode corresponding to the comoving Hubble diameter, $k_{aH} \equiv \pi aH$, with respect to the minimum and maximum resolvable modes in the box. 
Given a box length $L$, the smallest resolvable mode is given by the ``DC'' mode\footnote{Note that we include the volume correction $\sqrt{3}$ here, as this corresponds to the largest mode propagating \textit{diagonally} through the box.} $k_{DC} = 2\pi\sqrt{3}/L$ and the largest resolvable mode is given by the Nyquist mode, $k_{Nyq} = \pi N/L$.
For the super-horizon case, we validate the background dynamics.
For the horizon-crossing case, we demonstrate agreement with the Mukhanov-Sasaki equation \eqref{eqn:MS-soln-field} in the evolution of the tensor perturbations. 
This comparison allows us to validate our results against the linear solution, as well as to uncover a more faithful representation of semi-classical initial conditions.
We also present convergence tests on the constraints as well as the physical evolution of the first-order tensor power at a high mode, and we show preliminary calculations of the skewness and kurtosis of the polarisation fields.


\subsection{Super-horizon validation}
\label{subsec:sup-horizon}

\begin{figure}
    \centering
    \includegraphics[width=1.0\linewidth]{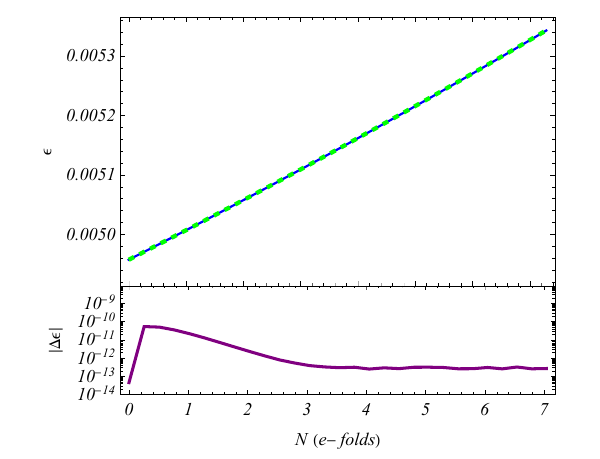}
    \caption{The first slow-roll parameter evolution for the super horizon case, extracted directly from GRChombo (solid blue) and from the Friedman equations (dashed green) using the same background initialisation. The difference is shown in purple.}
    \label{fig:srp1-sup}
\end{figure}

The super-horizon run initially satisfies $k_{Nyq} \sim k_{aH}$, such that all modes are initialised above the horizon scale. 
Given the model choice described in Sec.~\ref{subsec:ic-construction}, $k_{aH} \approx 2.57\cdot 10^{-3}\ m_{pl}$.
We choose to use $N=128$ grid points as the standard resolution, and therefore choose $L=16 \cdot 10^4 \, m_{pl}^{-1}$, which gives a mode range
\begin{eqnarray*}
    k_{DC} &\approx& 6.80\cdot 10^{-5} \ m_{pl}\\
    k_{Nyq} &\approx& 2.51\cdot 10^{-3}\ m_{pl}.
\end{eqnarray*}

Fig.~\ref{fig:sup-horizon-evo} shows the phases of evolution that the super-horizon run passes through, characterised by the location of the co-moving $k_{aH}$ with respect to $k_{Nyq}$.
We use the following parameters for the window function:
\begin{eqnarray*}
    k_* &=& 3/4\ k_{Nyq} \approx 1.88\cdot 10^{-3}\  m_{pl}\\
    \Delta_k &=& L/30 \approx 5333\ m_{pl}^{-1}.
\end{eqnarray*}
However, we do note that in this case, as dynamics are extremely suppressed on super-horizon scales, we expect that the window will not be entirely necessary in preserving the stability of the evolution.
We include it here in order to make contact with the horizon-crossing results presented later.

\begin{figure*}
    \begin{tikzpicture}
        \node at (0,0) {\includegraphics[width=0.48\linewidth]{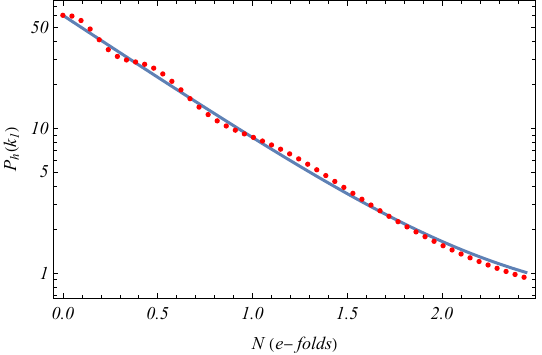}};
        \node at (-2.3, -1) {(a)};
    \end{tikzpicture}
    \hfill
    \begin{tikzpicture}
        \node at (0,0) {\includegraphics[width=0.48\linewidth]{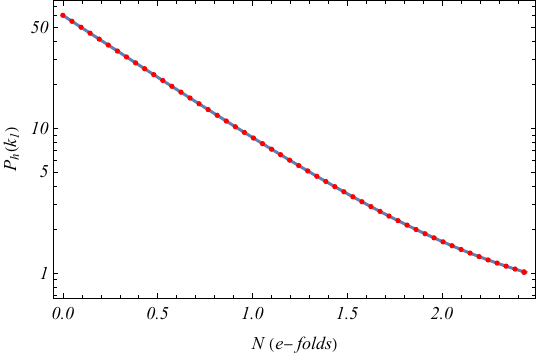}};
        \node at (-2.3, -1) {(b)};
    \end{tikzpicture}
    \caption{The evolution of the power $h_+^*h_+$ at a single mode $k_1 = 4\pi/L$ where $L=3000\ m_{pl}^{-1}$. This test was performed in GRChombo using the ``decoherent'' (a) and mode function (b) initialisation methods. Both GRChombo data sets (red, dotted) are compared against the Mukhanov-Sasaki solution (blue, solid) solved using the mode function initialisation method at $k_1$. This mode remains sub-horizon until approximately e-fold 2.}
    \label{fig:sqrt-vs-mode-fn-ps-evo}
\end{figure*}

We choose a background field which is described by the initial conditions laid out in Sec.~\ref{subsec:cpt}. 
We extract background quantities as box averages of the corresponding BSSN quantities,
\begin{equation}
    \label{eqn:H-a-from-chombo}
    H \approx -\frac{1}{3} \bar{K}, \ \ \ a \approx \frac{1}{\sqrt{\bar{\chi}}},
\end{equation}
however we do note some issues with this procedure recently have been brought forward in the literature, for instance in Ref.~\cite{aurrekoetxeaCosmologyUsingNumerical2024}. 
We suspect that this procedure is appropriate in our case, as we do not expect perturbations to become large in amplitude, and thus the spacetime should remain roughly homogeneous on the scale of the box.
We write the first slow-roll parameter under the slow-roll approximation as
\begin{eqnarray}
    \label{eqn:srp1-grchombo}
    \epsilon &\approx& \frac{3/2\dot{\bar{\phi}}^2}{\bar{\rho}},
\end{eqnarray}
since we can obtain $\dot{\bar{\phi}}$ and $\bar{\rho}$ easily from GRChombo.

Fig.~\ref{fig:srp1-sup} shows a comparison in the evolution of the first slow-roll parameter between GRChombo and the Friedman equations. 
We note that these two evolution schemes agree to a high precision, as expected from our model choice. 
We also observe that the first slow-roll parameter remains small ($\sim 10^{-3}$) over a large dynamic range (7 e-folds), which will be sufficient to capture full horizon-crossing behaviour. 


\subsection{Single mode validation}
\label{subsec:single-mode}

We now turn to the evolution of tensoral perturbations in this cosmological spacetime. 
In order to ensure a thorough match between our results in GRChombo and those produced by the Mukhanov-Sasaki equation, we performed a series of ``single mode'' tests, where we initialised only one plane waves in the box, and matched the power of this mode though the course of the simulation with the prediction from linear theory.\footnote{Note that this is not the power \textit{spectrum}, as the power spectrum in this case is not continuous, but rather a delta function in Fourier space.}
We performed this test using multiple methods of initialisation, some of which have been employed in the literature for the study of semi-classical perturbations in the early Universe. 

We present the most insightful of these comparisons in Fig.~\ref{fig:sqrt-vs-mode-fn-ps-evo}. 
In this test we compared the power evolution for two initialisation methods, which we will call the decoherent method (a), and the mode function method (b), for short. 
The decoherent method initialises the magnitude of each point in Fourier space in modulus/argument form as
\begin{equation}
    \label{sqrt-ps-init}
    h_s(k) = \sqrt{P_{h_s}(k)}e^{i\theta_s},
\end{equation}
where $P_{h_s}(k)$ is the power spectrum given by Eqn.~\eqref{eqn:field-spec}, and $\theta_s$ is a stochastic phase drawn from ${\cal U}(0, 2\pi)$. For this example we use the standard velocity corresponding to a free wave:
\begin{align*}
    \dot{h}_s(k) &= -ik h_s(k).
\end{align*}

The mode function method initialises the real and imaginary parts of the field exactly as given by Eqn.~\eqref{eqn:MS-soln-field} and \eqref{eqn:MS-soln-velocity}.
Transforming these functions into modulus/argument form, we find this method follows the structure
\begin{equation}
    \label{eqn:mode-fn-init}
    h_s(k) = \sqrt{P_{h_s}(k)}e^{i(\theta_{MS}(k)+\theta_s)}
\end{equation}
where we note the addition of a $k$-dependent phase coming directly from the Mukhanov-Sasaki solution.\footnote{The functional form of $\theta_{MS}$ is given in Apdx.~\ref{apdx:mod-arg-decomp}}
This $\theta_{MS}$ variable introduces a phase offset between the real and imaginary components \textit{at a particular point} in Fourier space.
However, it's inclusion does not require that separate points in Fourier space be phase-coherent, and the standard procedure of assigning a random phase to each point in Fourier space can still be used in addition to this variable.
The mode function for $\dot{h}_s$ is similarly decomposed, where as in the decoherent method we use the same random draw to set the stochastic elements of the velocity field and the position field at each point.

We note that the decoherent method is similar to the method commonly used in lattice studies of inflation, for example in Refs.~\cite{adsheadGaugePreheatingFull2024, figueroaArtSimulatingEarly2021,aurrekoetxeaOscillonFormationInflationary2023}. In these works, the magnitude of $h_s$ is set using a Rayleigh distribution centred on the power spectrum, as described in Sec.~\ref{subsec:ic-construction}.
We neglect this aspect of the stochastic realisation, as it only serves to introduce a difference in the overall scale of the GRChombo and Mukhanov-Sasaki solutions. We note, however, that the inclusion of the Rayleigh draw does not change our conclusion. 

As shown in Fig.~\ref{fig:sqrt-vs-mode-fn-ps-evo}, we recover the smooth decay of the power expected in the sub-horizon regime \textit{only} for the mode function method; or, in other words, when we take into account the k-dependent phase difference between the real and imaginary parts of both the initial field and velocity profile. 
This is because the real and imaginary components are both oscillatory in time, and so only with the correct relative phase will they add in quadrature to produce a smooth power spectrum evolution.
The oscillations in the decoherent method power evolution appear when this relative phase is not correctly assigned on the initial slice.


\subsection{Horizon-crossing validation}
\label{subsec:horizon-crossing}

The horizon-crossing case begins where $k \geq k_{aH}$ for all resolvable $k$, meaning that all modes start off dynamical. 
The Hubble diameter then moves into the box, freezing out modes from the largest to the smallest. 
Horizon crossing ends where $k < k_{aH}$ for all resolvable $k$.

We choose $N=128$ and $L=3\cdot 10^{3}\ m_{pl}^{-1}$ as the standard case.
This gives a mode range of 
\begin{eqnarray*}
    k_{DC} &\approx& 3.63\cdot 10^{-3}\ m_{pl}\\
    k_{Nyq} &\approx& 0.134 \ m_{pl}.
\end{eqnarray*}

\begin{figure}
    \centering
    \includegraphics[width=1.0\linewidth]{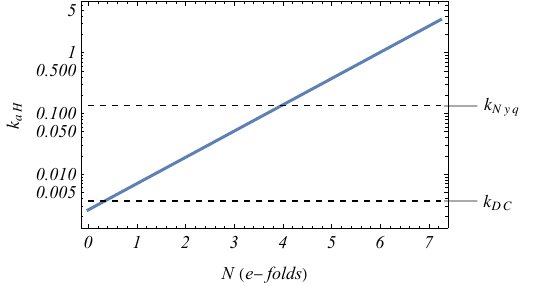}
    \caption{The Hubble mode $k_{aH}$ as a function of e-folds, for the horizon crossing run. The dotted black lines correspond to the size of the lowest resolvable mode, $k_{DC}$, and the highest resolvable mode, $k_{Nyq}$. }
    \label{fig:horizon-evo}
\end{figure}

Fig.~\ref{fig:horizon-evo} demonstrates these phases, in the same style as Fig.~\ref{fig:sup-horizon-evo}. 
Horizon crossing begins at about e-fold 0.5, and lasts until about e-fold 4.
Thus we should expect to see oscillatory behaviour in the polarisation fields initially, and then a gradual transition to a frozen state, completed by e-fold 4.

We use the same parameters for the window function as were used in the super-horizon case, taking into account the change in $L$:
\begin{eqnarray*}
    k_* &=& 3/4\ k_{Nyq} \approx 0.101 \  m_{pl}\\
    \Delta_k &=& L/30 = 100 \ m_{pl}^{-1}
\end{eqnarray*}
This produces the initial spectral shape shown in the blue line of Fig.~\ref{fig:spectrum-evolution}.
We set $A_{ij}$ according to the spectrum found in Sec.~\ref{subsec:ic-construction}, and apply the same window function to this spectrum. 

\begin{figure*}
    \begin{tikzpicture}
        \node at (0,0) {\includegraphics[width=0.48\linewidth]{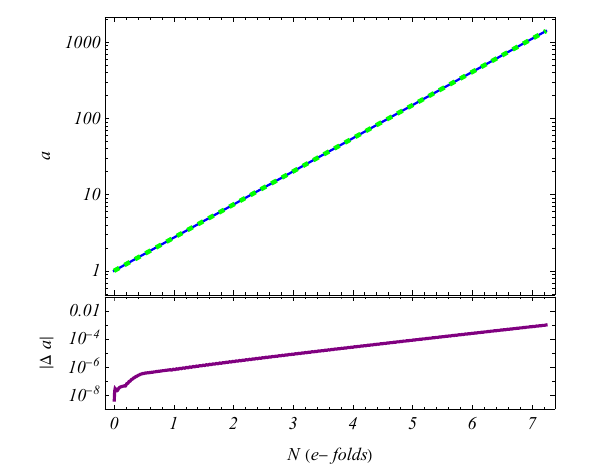}};
        \node at (-2.2, 2.5) {(a)};
    \end{tikzpicture}
    \hfill
    \begin{tikzpicture}
        \node at (0,0) {\includegraphics[width=0.48\linewidth]{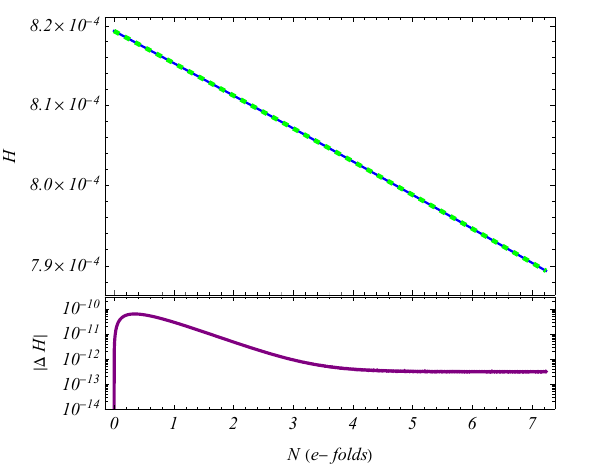}};
        \node at (3.0, 2.5) {(b)};
    \end{tikzpicture}
    \caption{The scale factor (a) and Hubble parameter (b) evolution in the horizon-crossing case, as calculated by GRChombo (blue, solid) and by the Friedman equations (green, dashed) for the same initial conditions. The differences between these solutions are shown in purple.}
    \label{fig:background-evo}
\end{figure*}

Fig.~\ref{fig:background-evo} shows a comparison between $a$ and $H$ as calculated by GRChombo, and as calculated by the Friedman equations with the same initial conditions.
We note that for all of these quantities, the error remains small over the course of the simulation, reflecting the fact that we are able to successfully recover the background solution, even in the dynamical regime.

\subsubsection{Recovery of spectral turnover}

We wish to match our solution for $h_s(k)$ to the solution of the Mukhanov-Sasaki equation, Eqn.~\eqref{eqn:h-cosmic-kg}, as we have chosen our model such that the slow-roll parameters are small. 
In order to accomplish this, we evolved the Mukhanov-Sasaki equation using the initial spectrum given in Eqn.~\eqref{eqn:MS-soln-field}. 
We do \textit{not} apply randomisation to this spectrum in the linearised case, but we do apply the same window function as is used in GRChombo.
We note the random draw for the modulus in the GRF formalism can cause disagreement with the results of Mukhanov-Sasaki, particularly where only a few modes are examined. 
However, we choose to compare the isotropic power spectrum from GRChombo with the results of Mukhanov-Sasaki along the same range of modes, in which case the power from a large number of modes with the same $|\textbf{k}|$ is integrated.
We find that in all but the first bin (where cosmic variance is more powerful) this average is sufficiently well-sampled that the we can recover the magnitude given by Mukhanov-Sasaki prediction well.

\begin{figure}
    \centering
    \includegraphics[width=1.0\linewidth]{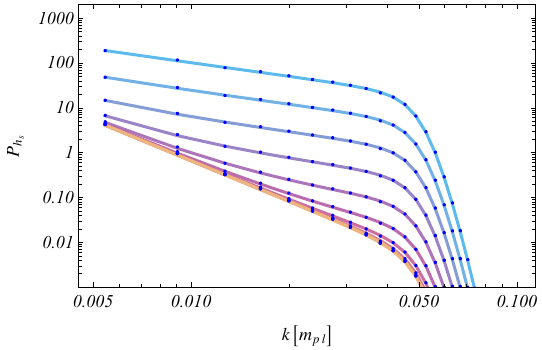}
    \caption{Evolution of the power spectrum of $h_+(k)$, from e-fold 0 (light blue) to e-fold 7 (yellow). The binned power spectrum extracted from GRChombo is shown in dark blue dots, and the linearised spectrum evolution is shown by the solid lines.}
    \label{fig:spectrum-evolution}
\end{figure}

Fig.~\ref{fig:spectrum-evolution} compares the evolution of the spectrum between the Mukhanov-Sasaki solution and the numerical solution.
The solid lines represent the Mukhanov-Sasaki power spectrum at different time slices, from blue on the initial slice to yellow on the final slice.
Eleven slices are shown in total, sampled evenly between e-fold 0 and 7, meaning that one spectrum is shown for every $\sim 0.7$ e-folds.
The dark blue dots represent the isotropic binned power spectrum extracted from GRChombo at the same time slice.
We include all bins between $k_2 = 2\cdot\frac{2\pi}{L}$ and $k_{30}=30\cdot\frac{2\pi}{L}$, after which the window function dominates the signal.
The initial and final spectra shown in particular demonstrate the recovery of the correct sub-horizon and super-horizon scaling with $k$, as described by Eqn.~\eqref{eqn:MS-power-spectrum} .
The power spectrum derived from GRChombo matches the Mukhanov-Sasaki power spectrum quite well for all modes shown, which are the modes best protected from the numerical issues. 

\begin{figure*}
    \begin{tikzpicture}
        \node at (0,0) {\includegraphics[width=0.48\linewidth]{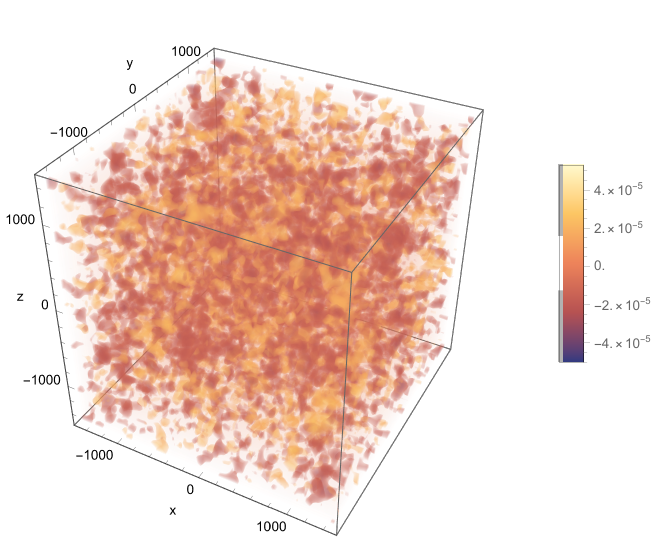}};
        \node at (-3.2, -3.2) {(a)};
    \end{tikzpicture}
    \hfill
    \begin{tikzpicture}
        \node at (0,0) {\includegraphics[width=0.48\linewidth]{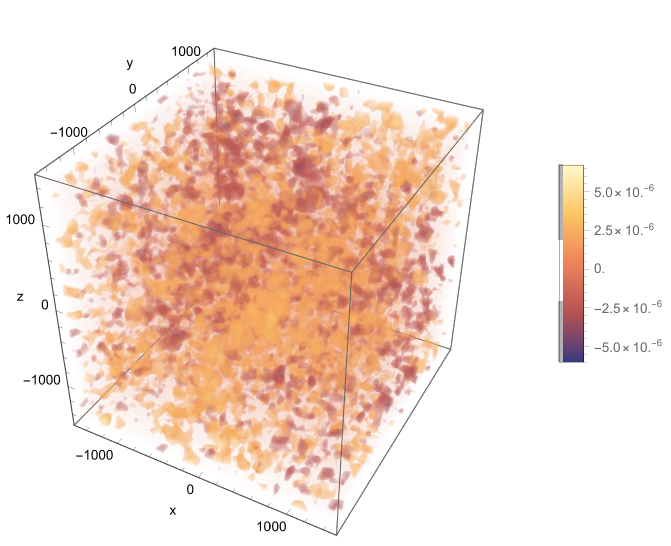}};
        \node at (-3.2, -3.2) {(b)};
    \end{tikzpicture}
    \vfill
    \begin{tikzpicture}
        \node at (0,0) {\includegraphics[width=0.48\linewidth]{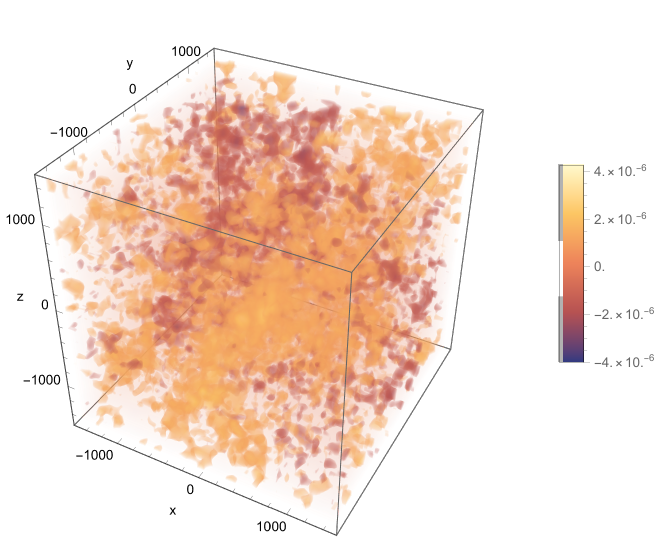}};
        \node at (-3.2, -3.2) {(c)};
    \end{tikzpicture}
    \hfill
    \begin{tikzpicture}
        \node at (0,0) {\includegraphics[width=0.48\linewidth]{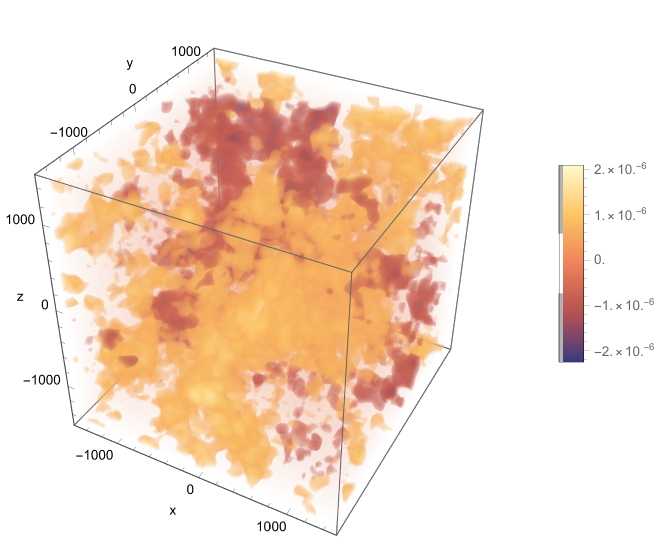}};
        \node at (-3.2, -3.2) {(d)};
    \end{tikzpicture}
    \caption{Density plots of the amplitude of $h_+(\textbf{x})$, extracted from GRChombo at four points in time: the initial slice (a), the final slice (d), and two slices during horizon crossing: where $N = 2.1$ (b), and where $N = 2.9$ (c). We have allowed the colour scale to vary in each plot, in order to highlight the change of structure which occurs in these polarisation fields separately from the overall scale change. Note that in our units, length is measured in units of $m_{pl}^{-1}$.}
    \label{fig:box-plots}
\end{figure*}
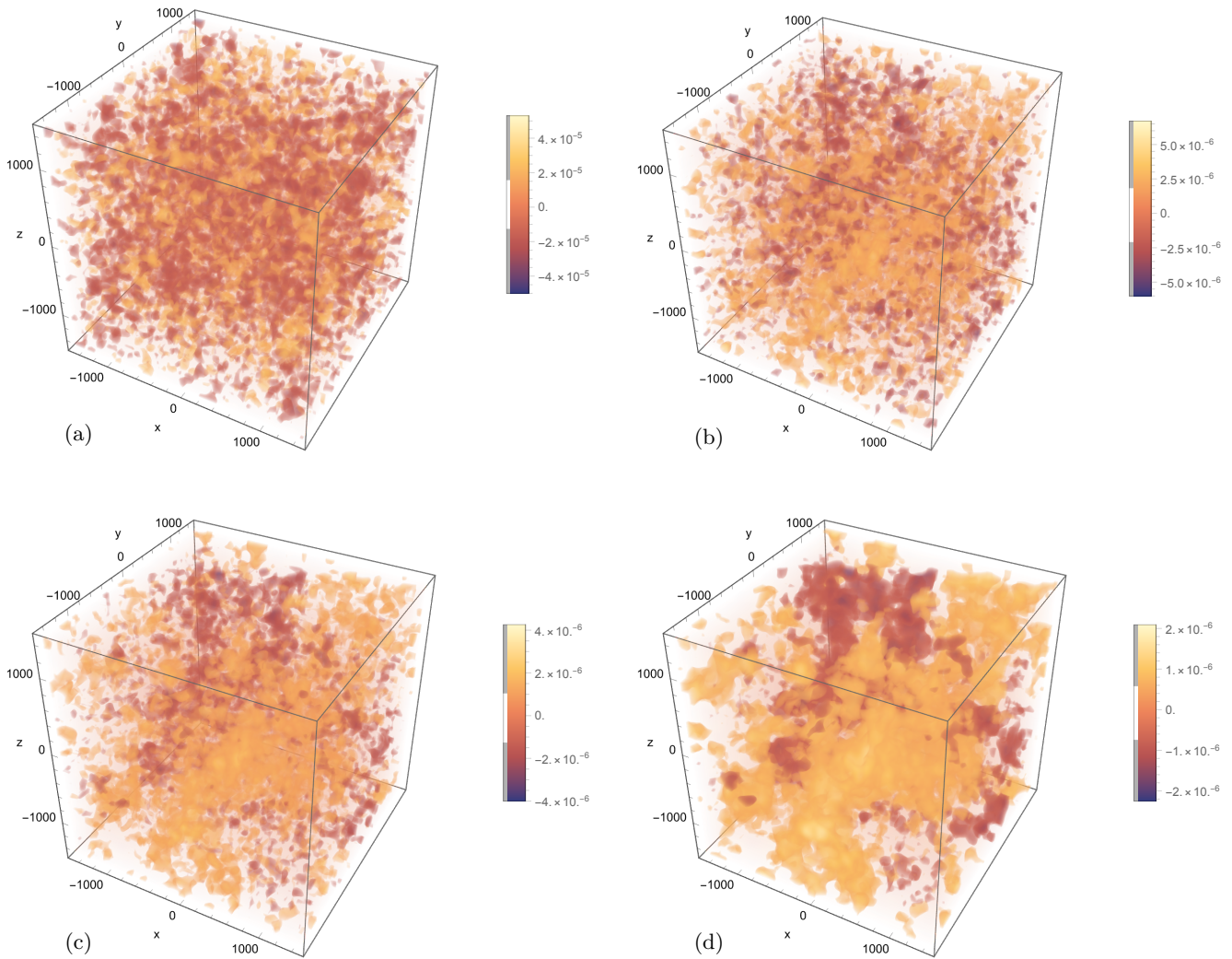

Fig.~\ref{fig:box-plots} shows the configuration-space representation of this spectral phenomenon. 
Here we present four density plots showing the amplitude of $h_+(\textbf{x})$ through the course of the simulation.
We note the emergence of structure at low modes, corresponding to the transition to a $1/k^3$ scaling.
We find that this transition occurs quite obviously between approximately e-folds 2 and 3.5, a period where many of the intermediate modes are frozen out.
As the skewness and kurtosis parameters are calculated directly from the field depicted here, the transition that this figure demonstrates will be key to understanding the evolution of the skewness and kurtosis, presented in the final section.


\subsection{Constraints, stability and convergence}
\label{subsec:convergence-and-constraints}

For a perfect solution to Einstein's equations, ${\cal H} = {\cal M}_i = 0$. 
However, numerical solutions to Einstein's equations can never perfectly replicate the continuum solution, and so some violation of these constraint equations is expected.
Stable NR simulations are characterised by constraint variables which are initially small, compared to some relevant physical scale, and which do not grow with time at any point during the simulation.
This ensures that the numerical solution stays close to the continuum solution surface.

In order for a realistic numerical solution to be found, the initial data must satisfy the constraints to an acceptable degree. 
Where the initial curvature of the metric may be large, the initial data can be found by solving the constraints iteratively for some initial ansatz metric and matter configuration.\footnote{The recently-developed CTTK solver can accomplish this task for GRChombo simulations, see Ref.~\cite{aurrekoetxeaCTTKNewMethod2023} for further details.} 
However, as we plan to stay within perturbation theory for the sake of the following validation, we do not expect large initial curvature to be present.
As illustrated at the end of Sec.~\ref{subsec:cpt-bssn-dict}, as long as $h_{ij}$ and $\dot{h}_{ij}$ remain transverse, traceless, and small relative to the background, we expect that the constraints will be satisfied at first order, meaning that they should have a magnitude on the order of $\sigma_h^2$ where $\sigma_h$ measures the size of the tensor perturbations. 
We choose the sample standard deviation of the polarisation fields as a measure of their amplitude on the lattice, and we can approximate this measure as 
\begin{equation*}
    \sigma_h \approx \sqrt{\frac{1}{N^3}\sum_{i,j,k} h_s^2(i,j,k)},
\end{equation*}
since the mean of the $h_s$ fields in configuration space lies at machine precision.

\begin{figure*}
    \begin{tikzpicture}
        \node at (0,0) {\includegraphics[width=0.48\linewidth]{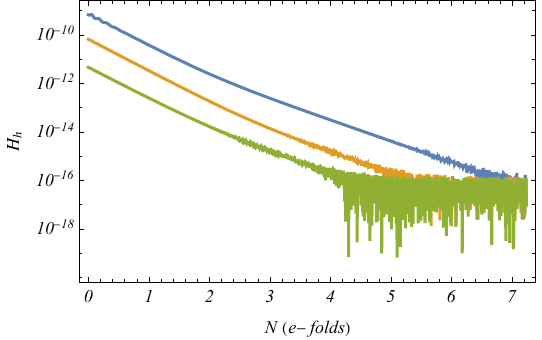}};
        \node at (-2.5, -1.3) {(a)};
    \end{tikzpicture}
    \hfill
    \begin{tikzpicture}
        \node at (0,0) {\includegraphics[width=0.48\linewidth]{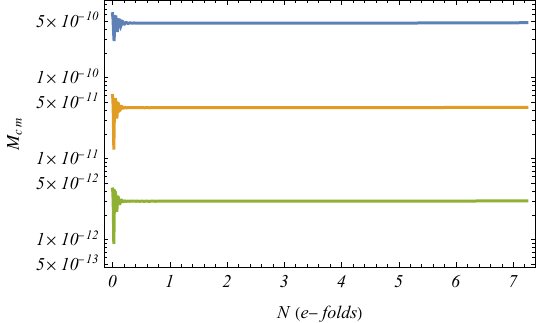}};
        \node at (-2.0, -1.2) {(b)};
    \end{tikzpicture}
    \caption{Sample averages of the constraints in the horizon-crossing case for three box resolutions: $N=64$ (blue), $N=128$ (orange) and $N=256$ (green). Panel (a) shows the residual Hamiltonian ${\cal H}_h$, and panel (b) shows the co-moving momentum constraint, ${\cal M}_{cm}$. }
    \label{fig:constraint-evo}
\end{figure*}

NR simulations also typically utilise some form of dynamical constraint damping.
However, as detailed in Appendix \ref{apdx:ko-dissipation}, dynamical constraint damping can have a strong effect on the spectrum, which could introduce non-physical signals in our analysis.
Additionally, de-Sitter space is naturally damping, in that it suppresses higher-order terms in the constraints\footnote{In other words, geodesics naturally separate in de-Sitter, and thus their crossing becomes much less likely (Ref~\cite{aurrekoetxeaCosmologyUsingNumerical2024}).} by a factor of $a^2$.
Therefore we have disabled all forms of dynamical dissipation for the purpose of validation. 
This means that our simulation could be more prone to leaving the constrained surface of classical solutions.
However, a robust reproduction of linear results in this case would provide greater evidence that numerical simulations can be used to go beyond the linear regime safely, as they may not require damping to stay near the true cosmological solution.

\subsubsection{Constraint characterisation}

Since the BSSN constraint variables ${\cal H},{\cal M}_i$ become zero on the solution surface, rarely is there an obvious method to compare their magnitude to some relevant physical quantity. 
In fact, the appropriate comparison for these constraints will often differ depending on the system one is attempting to solve. 
In arriving at this comparison scheme, it is important to consider what values these constraints can possibly take on.
The Hamiltonian constraint could take on positive or negative values, depending on the arrangement of the perturbation field. 
The momentum constraint, however, is characterised in GRChombo by its vector magnitude,
\begin{equation*}
    {\cal M} \equiv |{\cal M}_i| = \sqrt{\sum_i M_i^2}.
\end{equation*}
and so is always positive. 

GRChombo defines two alternative constraint measures which can be used to normalise ${\cal H}$ and ${\cal M}$. 
These are the ``absolute'' constraint variables, defined to be the sum of the absolute values of all terms in the constraint equations, and are given by
\begin{eqnarray*}
    {\cal H}_{abs} &\equiv& |R + \frac{2}{3}K^2 + \tilde{A}_{ij}\tilde{A}^{ij} + 16\pi G \rho|\\
    {\cal M}_{abs} &\equiv& |\tilde{\gamma}^{kl}\left( \partial_k \tilde{A}_{il} + 2\tilde{\Gamma}^m_{l(i}\tilde{A}_{k)m} + 3 \tilde{A}_{ik} \frac{\partial_l \chi}{2\chi} \right)\\
    & & \hspace{0.3\linewidth} + \frac{2}{3} \partial_i K + 8\pi G S_i|.
\end{eqnarray*}
Note that ${\cal H}_{abs} \neq |{\cal H}|$, as $|{\cal H}|$ allows for the cancellation between different terms at each point, whereas ${\cal H}_{abs}$ does not. 

Under the CPT-BSSN correspondence, we note that the mean of ${\cal H}_{abs}$ should measure twice the energy density.
This energy density will be made up of a background component, captured by the Friedman equation, and a perturbative part arising from the presence of tensor perturbations.
Where this gravitational wave energy density is fully accounted for in the initial matter sector (see discussion in Ref.~\cite{otaCovariantTransversetracelessProjection2022}), the constraints may be satisfied beyond second order.
However, in our case, we only wish to show that these tensors contribute to the Hamiltonian constraint violation at second order.
Thus we define a diagnostic for the Hamiltonian constraint violation due to the presence of tensor perturbations as
\begin{equation}
    {\cal H}_{h} \equiv {\cal H}_{abs} - 12H^2.
\end{equation}
This definition eliminates the homogeneous contribution to ${\cal H}_{abs},$ leaving only the tensoral component, which will go like $\sigma_h^2$ if the constraints are satisfied to second order.

We note that the momentum constraint does not have a contribution from the background.
Thus the raw momentum constraint should go like $\sigma_h^2$.
We note also that a factor of $1/a^2$ multiplies the second-order tensoral contributions, which arises from the background term in the inverse spatial metric. 
Thus we define the \textit{co-moving} momentum constraints as 
\begin{eqnarray*}
    {\cal M}_{cm} &\equiv& a^2 {\cal M}.
\end{eqnarray*}
These measure the constraint violation on co-moving scales.

Fig.~\ref{fig:constraint-evo} shows the evolution of the sample mean of ${\cal H}_{h}$ (panel a) and ${\cal M}_{cm}$ (panel b) as a function of e-folds, in the horizon-crossing case. 
These runs were performed with $L=3000\ m_{pl}^{-1}$ and $H_0 \approx 8.2\cdot 10^{-4}\ m_{pl}$, which produces an initial tensor polarisation field with $\sigma_h \sim 10^{-5}$.
We note that even for the coarsest resolution, the initial Hamiltonian and momentum constraint violations lie within an order of magnitude of $\sigma_h^2$. 
For each resolution, the constraints continue to either damp away with $a^2$, in the case of the raw constraint measure, or to remain constant, in the case of the co-moving constraint measure, and at no point do the constraints show consistent growth.
Thus our simulations satisfy this check of numerical stability.

We point out that since the raw constraints scale with $a^{-2}$, they may decrease sufficiently quickly such that they will reach floating-point precision within the dynamical range of the simulation.
This is demonstrated in the evolution of ${\cal H}_h$, where we note that in modern C++ the floating-point precision for double-type variables is $10^{-15}$.
This effect may produce an apparent amplification in the constraints, particularly in the co-moving constraints, that is \textit{not} due to an instability in the evolution scheme itself.
Parallel analysis of the raw and co-moving constraints is therefore recommended, in order to avoid this pitfall.

\subsubsection{Convergence tests}

We have performed convergence tests on the constraint evolution, and on the evolution of the power spectrum, at three resolutions: $N_c = 64$, $N_m = 128$ and $N_f = 256$.  
The initial conditions for these runs were were coarse-grained off of the finest resolution, so that each run represent the same draw from the same statistical distribution.\footnote{This means, for instance, that the initial conditions for $N_c$ were found by generating the initial conditions for $N_f$, and placing every 4th point on the GRChombo grid.}
In each case we used a window function characterised by
\begin{align*}
    k_* = \pi N_c/L\cdot 3/4,
\end{align*}
such that increasing resolution did not add any new modes with significant power, but only increased the accuracy with which our initial range of modes was resolved.
Thus the results of these tests should converge purely on the classical dynamics, they should not contain spurious noise due to the inherent stochasticity of the initial conditions, and any numerical error due to the evolution of modes close to the Nyquist frequency should be increasingly suppressed.

\begin{figure}
    \centering
    \includegraphics[width=1.0\linewidth]{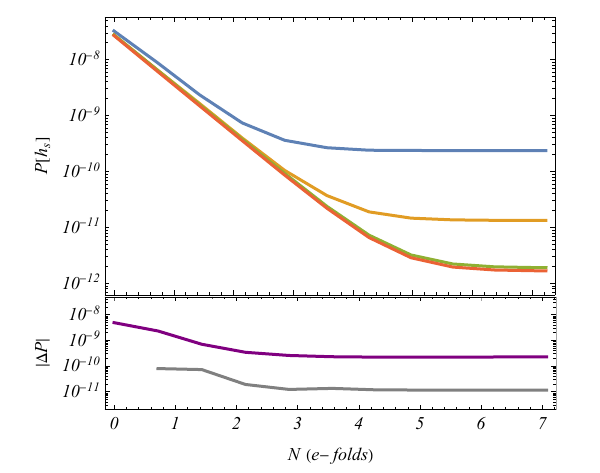}
    \caption{The evolution of the power in $h_+(k)$ at the isotropic bin $k_{28}$, for $N_c$ (blue), $N_m$ (orange) and $N_f$ (green). This is compared with the prediction from Mukhanov-Sasaki, solved at $k_{28}$ (red). The differences $|N_m-N_c|$ (purple) and $|N_f-N_m|$ (grey) are shown in the inset.}
    \label{fig:mf-convergence}
\end{figure}

Fig.~\ref{fig:mf-convergence} shows convergence on the evolution of one bin in the power spectrum. 
We choose a high bin for this test, $k_{28} = 28\sqrt{3}\cdot 2\pi/L$, as high modes are less numerically stable and so convergent behaviour here can be clearer.
Note that for each run, $k_{28}$ represents the evolution of a mode at approximately $88\%$ ($N_c$), $44\%$ ($N_m$) and $22\%$ ($N_f$) of the Nyquist frequency.
As shown, we recover strong visual convergence towards the Mukhanov-Sasaki solution, even at this high mode.
We show that our solution converges at second-order with the residuals between each curve.
We expect second-order convergence in time, as in GRChombo the time-evolution algorithm is second-order accurate (Ref.~\cite{radiaLessonsAdaptiveMesh2022}).
However, examining the constraint evolution in Fig.~\ref{fig:constraint-evo}, we see roughly one order of magnitude decrease in the initial value of these constraints with increasing $N$, suggesting that we can achieve first-order convergence in the constraints.


\subsection{First look at non-Gaussianity}
\label{subsec:non-gauss}

\begin{figure}
    \centering
    \includegraphics[width=1.0\linewidth]{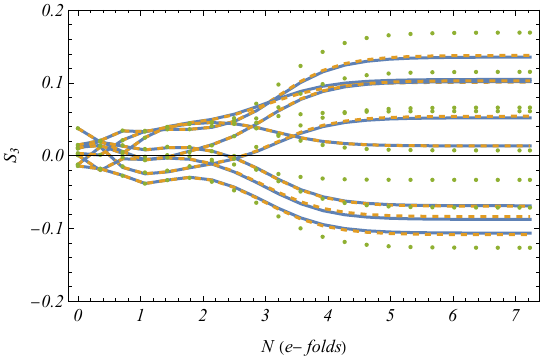}
    \caption{The skewness of $h_+(\textbf{x})$ and $h_{\times}(\textbf{x})$, as a function of e-folds, extracted from GRChombo for $N=128$. Here we show the skew evolution for four runs with different random seeds, for the amplitude values $A=1$ (blue, solid), $A=100$ (orange, dashed) and $A=1000$ (green, dotted).}
    \label{fig:skew-paths}
\end{figure}

A key goal of this project is the explore non-Gaussianity and, specifically, tensoral bispectra and cross-spectra as cosmological observables which can distinguish between the predictions of a wide variety of inflationary scenarios. In practice, this will involve exploiting the efficient MODAL bispectrum estimation pipeline (Ref.~\cite{fergussonCMBBispectrum2012,fergussonPrimordialNonGaussianityCMB2007}), where the separable methodology has been applied to constrain many inflation models using Planck satellite data \cite{Planck:2018jri}; in this case, the 3D implementation will be applied first to extract primordial bispectra \cite{Fergusson:2010ia,hung2019advancingmatterbispectrumestimation}.
However, as the purpose of the present paper is to implement the tensor evolution equations with gravitational back-reaction, together with appropriate initial conditions, we focus on validating the two-point correlator dynamics in perturbative regimes and only give some indicative results of the emergence of small higher-point correlators. 

Here, we will identify the presence of the bispectrum and trispectrum by simply calculating the skew and kurtosis parameters of each polarisation field in configuration space. The skew parameter is defined as 
\begin{equation*}
    \mu_3 \equiv \frac{\kappa_3}{\kappa_2^{3/2}} = \frac{E[(h_s(\textbf{x}) - \bar{h}_s)^3]}{E[(h_s(\textbf{x}) - \bar{h}_s)^2]^{3/2}}
\end{equation*}
where $\kappa_i$ is the $i^{th}$ cumulant of the polarisation field. The kurtosis is defined as
\begin{equation*}
    \mu_4 \equiv \frac{\kappa_4}{\kappa_2^4} = \frac{E[(h_s(\textbf{x}) - \bar{h}_s)^4]}{E[(h_s(\textbf{x}) - \bar{h}_s)^2]^{2}}.
\end{equation*}
The expectation values are taken to be sample averages on the whole grid. 

\begin{figure*}
    \begin{tikzpicture}
        \node at (0,0) {\includegraphics[width=0.48\linewidth]{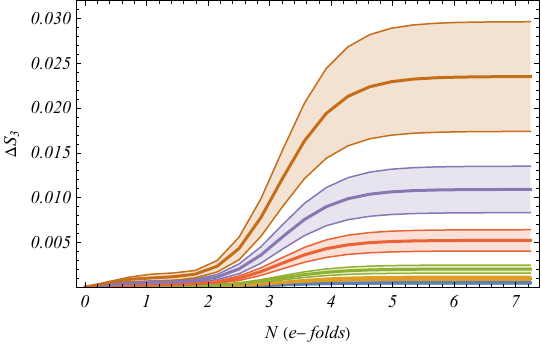}};
        \node at (-2.4, 2.1) {(a)};
    \end{tikzpicture}
    \hfill
    \begin{tikzpicture}
        \node at (0,0) {\includegraphics[width=0.48\linewidth]{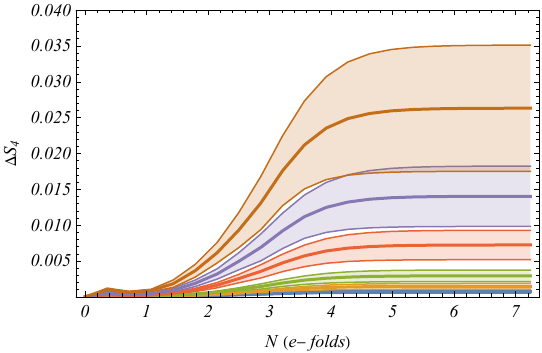}};
        \node at (-2.4, 2.1) {(b)};
    \end{tikzpicture}
    \caption{The difference between the skew (a) and kurtosis (b) paths as a function of e-folds, for $A=25$ (blue), 50 (orange), 100 (green), 250 (red), 500 (purple) and 1000 (brown), where in each case the difference is taken with respect to the $A=1$ solution. For each data set, the mean (solid) and first standard deviation (filled bands) over random field configurations are shown.}
    \label{fig:stats-diffs}
\end{figure*}

We performed a series of horizon-crossing GRChombo simulations with different random seed values, while varying the initial tensoral amplitude $A$, increasing from standard perturbative values with $A=1$ in a realistic inflation scenario  (giving $\sigma_h\sim 10^{-5}$ initially), up to much larger quasi-nonlinear value $A=1000$ (giving $\sigma_h\sim 10^{-2}$ initially). 
We note that even for a random Gaussian distribution, there is always a stochastic ``shot noise'' contribution to the skewness and kurtosis, dependent on the finite sample size used in the numerical implementation. 
The kurtosis is also known to be biased, and this bias is accentuated for the relatively small grids investigated here. 
For this reason, we will analyse the small difference between the skewness and kurtosis as a function of increasing $A$, i.e.\ $\Delta S_3 \equiv |S_3(A)-S_3(1)|$. We define $\Delta S_4$ similarly. 
Fig.~\ref{fig:skew-paths} shows these difference measures, where each colour corresponds to a different value of $A$. 
These statistics are calculated from a collection of 8 independent samples of the skew and kurtosis trajectories, as the polarisation fields evolve independently in the linear regime.

We note a correspondence between the behaviour of both statistics and the expected behaviour of the polarisation fields, in that these statistics oscillate in the sub-horizon regime, grow during horizon-crossing and then freeze at a constant value afterwards. 
We note in particular that this apparent growth and the final freeze-out value occurs where the super-horizon and horizon-crossing runs overlap and at the same value for small amplitude $A$, suggesting that this process is largely caused by linear-order dynamics; that is, the linear evolution of the skewness by the time $t_f$ gives the same result as the initial conditions set at the time $t_f$ for the same random seed. 
Most of the growth in the stochastic skewness signal can be understood as due to the evolution of the power spectrum $P(k)$ from the Bunch-Davies $k^{-1}$ on sub-horizon scales transforming to the super-horizon $k^{-3}$ during inflation.  
The skewness corresponds to a summation over all the bispectrum from large to small physical scales (normalised to the same with respect to $P(k)^2$), where the sum is equivalent to an integration over the measure $k^2 dk$.  
This means that when the power spectrum is $k^{-1}$, there is more weight given to the highly populated large $k$ modes (with less ``shot noise''), whereas later with $k^{-3}$ there is greater weight at small $k$, enhancing the stochastic noise contribution; hence, in linear theory, the transition shown in Fig.~\ref{fig:skew-paths} from small to larger skewness is a result of the power spectrum being transformed at horizon-crossing.

The key signal revealing nonlinearity, therefore, is the subtle growth in the skewness shown in Fig.~\ref{fig:stats-diffs} that arises from increasing the amplitude $A$ of the initial tensor perturbations for an identical simulation with the same random seed. 
Here, there is a discernible difference in the skewness $\Delta S_3$ (and kurtosis $\Delta S_4$) above $A>100$ which grows further with increasing amplitude $A$. 
We note that the standard deviation $\sigma_h$ of both measures on the initial slice also increases. 
We conclude that this a nonlinear evolutionary effect, which does not occur for linear stochastic noise, and nor are these deviations as obvious for the power spectrum $P(k)$, which simply grows with the expected $A^2$.  
Finally, we note that for a time after horizon-crossing, the tensor amplitude remains ``frozen", indicating that it is a nonlinearly conserved quantity, analogous to the scalar curvature perturbation $\zeta$ which is nonlinearly conserved for single field inflation. 
Note that we will undertake a quantitative study of non-Gaussianity by estimating the tensoral bispectrum directly for a variety of inflationary scenarios in a future publication \cite{EF-EPS:Future}.

\section{Summary and future directions}
\label{sec:summary}
Primordial non-Gaussian signals are expected to provide unique insights in the effort to uncover the true nature of the inflationary epoch.
In order to fully understand how the coupling between different perturbations alters the resulting spectral profile, we should include terms at all orders in the equations of motion. 
Numerical relativity allows us to solve Einstein's equations entirely non-perturbatively, so this enable classical gravitational contributions at all orders to be included.
Nevertheless, inflation is expected to be a largely perturbative process, and while nonlinear signals may allow us to distinguish inflationary scenarios, these signals are expected to be small relative to the background.
This has necessitated the development of a new set of tools for the use of NR in cosmology, which are adapted to problems where nonlinear physics may play a relatively modest role and then a significant role at different points in the evolution.

In this work, we have presented a gauge-independent dictionary which translates between the perturbative description of inflation, and the fully-relativistic variables of NR.
We have used this dictionary with a straightforward gauge choice to develop a new version of the GRChombo code which evolves tensoral perturbations on an inflating background.
We have validated the dynamics of the inflationary background, in both the super-horizon and horizon-crossing case. For tensor perturbations  with given inflationary initial conditions, we have accurately matched the evolution of the two-point correlator or power spectrum to the results of the Mukhanov-Sasaki solution.
Thus, we have demonstrated that our NR simulation is able to match the evolution of gravitational waves during inflation predicted by linear theory. 
Furthermore, this investigation has presented a series of tests which can be used to validate the accuracy of our results, as well as other NR codes which evolve inflationary spacetimes.
Through these tests, we have uncovered the presence of a $k$-dependent phase in the Mukhanov-Sasaki solution which yields initial conditions for the lattice simulations on sub-horizon scales that more closely reproduce the power spectrum evolution from linear theory.

We have analysed the behaviour of the energy and momentum constraints for a perturbative gravitational wave background. We have found that these constraints are initially second-order in magnitude or smaller, and remain stable or diminish, depending on the particular constraint diagnostic used.
We have also demonstrated the expected second-order convergence of the tensor power evolution.
We then gave a first look into how our pipeline can be used to capture the emergence of non-Gaussian signals from higher-order gravitational couplings to the metric.
We calculated the amplitude of the bispectrum, encapsulated by the skewness parameter, for a series of runs with different random configurations, and measured the change in the amplitude of the bispectrum as we increased the amplitude of these perturbations.

We will use this pipeline to investigate the emergence of non-Gaussianity in models of inflation that exhibit strong nonlinear dynamics, such as ultra slow-roll phases (Ref.~\cite{launayStochasticInflationGeneral2024a}).
We also plan to combine our work with that of Ref.~\cite{launayStochasticInflationGeneral2024a}, in order to create an NR program which can evolve scalar and tensor perturbations together, incorporating their interaction fully into the equations of motion.
This will allow us to measure mixed bispectrum signals, such as $\langle\zeta\zeta\gamma\rangle$, which may be dominant over the tensor-only bispectrum in future observations.
Once this is complete, we will be able to combine our program with the MODAL pipeline to produce full bispectrum estimates which can be used to analyse next-generation CMB data.

\begin{acknowledgments}
EF would like to acknowledge financial support from the GAPSTI Foundation and an Isaac Newton Scholarship, both held within the STFC CDT in Data Intensive Science.  EPS acknowledges partial support from STFC Consolidated Grants ST/X000664/1 and ST/P000673/1 and from the CTC Intel oneAPI Centre.  This has also supported Dr Juliana Kwan who has provided expert programming guidance and developmental help with GRChombo/GRTeclyn.  The authors would particularly like to thank Yoann Launay for many insightful comments on stochastic inflation and Ulrich Sperhake for his expertise in numerical relativity, along with David Baker and Bowei Zhang for helpful discussions. We would also like to acknowledge the members of the GRTL community for maintenance and improvement of the GRChombo/GRTeclyn codebase, especially Katy Clough, Eugene Lim, Josu Aurrekoetxea, Miren Radia, and Amelia Drew.   Part of this work was
undertaken on the Cambridge CSD3 part of the STFC DiRAC HPC Facility (www.dirac.ac.uk)
funded by BEIS capital funding via STFC Capital Grants ST/P002307/1 and ST/R002452/1
and STFC Operations Grant ST/R00689X/1.
\end{acknowledgments}

\bibliography{bibliography}

\begin{thebibliography}{79}%
\makeatletter
\providecommand \@ifxundefined [1]{%
 \@ifx{#1\undefined}
}%
\providecommand \@ifnum [1]{%
 \ifnum #1\expandafter \@firstoftwo
 \else \expandafter \@secondoftwo
 \fi
}%
\providecommand \@ifx [1]{%
 \ifx #1\expandafter \@firstoftwo
 \else \expandafter \@secondoftwo
 \fi
}%
\providecommand \natexlab [1]{#1}%
\providecommand \enquote  [1]{``#1''}%
\providecommand \bibnamefont  [1]{#1}%
\providecommand \bibfnamefont [1]{#1}%
\providecommand \citenamefont [1]{#1}%
\providecommand \href@noop [0]{\@secondoftwo}%
\providecommand \href [0]{\begingroup \@sanitize@url \@href}%
\providecommand \@href[1]{\@@startlink{#1}\@@href}%
\providecommand \@@href[1]{\endgroup#1\@@endlink}%
\providecommand \@sanitize@url [0]{\catcode `\\12\catcode `\$12\catcode
  `\&12\catcode `\#12\catcode `\^12\catcode `\_12\catcode `\%12\relax}%
\providecommand \@@startlink[1]{}%
\providecommand \@@endlink[0]{}%
\providecommand \url  [0]{\begingroup\@sanitize@url \@url }%
\providecommand \@url [1]{\endgroup\@href {#1}{\urlprefix }}%
\providecommand \urlprefix  [0]{URL }%
\providecommand \Eprint [0]{\href }%
\providecommand \doibase [0]{https://doi.org/}%
\providecommand \selectlanguage [0]{\@gobble}%
\providecommand \bibinfo  [0]{\@secondoftwo}%
\providecommand \bibfield  [0]{\@secondoftwo}%
\providecommand \translation [1]{[#1]}%
\providecommand \BibitemOpen [0]{}%
\providecommand \bibitemStop [0]{}%
\providecommand \bibitemNoStop [0]{.\EOS\space}%
\providecommand \EOS [0]{\spacefactor3000\relax}%
\providecommand \BibitemShut  [1]{\csname bibitem#1\endcsname}%
\let\auto@bib@innerbib\@empty
\bibitem [{\citenamefont {Tsujikawa}\ \emph {et~al.}(2013)\citenamefont
  {Tsujikawa}, \citenamefont {Ohashi}, \citenamefont {Kuroyanagi},\ and\
  \citenamefont {De~Felice}}]{tsujikawaPlanckConstraintsSinglefield2013}%
  \BibitemOpen
  \bibfield  {author} {\bibinfo {author} {\bibfnamefont {S.}~\bibnamefont
  {Tsujikawa}}, \bibinfo {author} {\bibfnamefont {J.}~\bibnamefont {Ohashi}},
  \bibinfo {author} {\bibfnamefont {S.}~\bibnamefont {Kuroyanagi}},\ and\
  \bibinfo {author} {\bibfnamefont {A.}~\bibnamefont {De~Felice}},\ }\href
  {https://doi.org/10.1103/PhysRevD.88.023529} {\bibfield  {journal} {\bibinfo
  {journal} {Physical Review D}\ }\textbf {\bibinfo {volume} {88}},\ \bibinfo
  {pages} {023529} (\bibinfo {year} {2013})}\BibitemShut {NoStop}%
\bibitem [{\citenamefont {Martin}\ \emph {et~al.}(2014)\citenamefont {Martin},
  \citenamefont {Ringeval}, \citenamefont {Trotta},\ and\ \citenamefont
  {Vennin}}]{martinBestInflationaryModels2014}%
  \BibitemOpen
  \bibfield  {author} {\bibinfo {author} {\bibfnamefont {J.}~\bibnamefont
  {Martin}}, \bibinfo {author} {\bibfnamefont {C.}~\bibnamefont {Ringeval}},
  \bibinfo {author} {\bibfnamefont {R.}~\bibnamefont {Trotta}},\ and\ \bibinfo
  {author} {\bibfnamefont {V.}~\bibnamefont {Vennin}},\ }\href
  {https://doi.org/10.1088/1475-7516/2014/03/039} {\bibfield  {journal}
  {\bibinfo  {journal} {Journal of Cosmology and Astroparticle Physics}\
  }\textbf {\bibinfo {volume} {2014}}\bibfield  {number} {\bibinfo  {number} {
  (03)},\ \bibinfo {pages} {039}},\ }\Eprint {https://arxiv.org/abs/1312.3529}
  {arXiv:1312.3529 [astro-ph, physics:gr-qc, physics:hep-ph, physics:hep-th]}
  \BibitemShut {NoStop}%
\bibitem [{\citenamefont {{BICEP/Keck
  Collaboration}}(2021)}]{bicep/keckcollaborationImprovedConstraintsPrimordial2021}%
  \BibitemOpen
  \bibfield  {author} {\bibinfo {author} {\bibnamefont {{BICEP/Keck
  Collaboration}}},\ }\href {https://doi.org/10.1103/PhysRevLett.127.151301}
  {\bibfield  {journal} {\bibinfo  {journal} {Physical Review Letters}\
  }\textbf {\bibinfo {volume} {127}},\ \bibinfo {pages} {151301} (\bibinfo
  {year} {2021})}\BibitemShut {NoStop}%
\bibitem [{\citenamefont {Starobinski{\v
  i}}(1979)}]{starobinskiiSpectrumRelictGravitational1979}%
  \BibitemOpen
  \bibfield  {author} {\bibinfo {author} {\bibfnamefont {A.~A.}\ \bibnamefont
  {Starobinski{\v i}}},\ }\href@noop {} {\bibfield  {journal} {\bibinfo
  {journal} {Soviet Journal of Experimental and Theoretical Physics Letters}\
  }\textbf {\bibinfo {volume} {30}},\ \bibinfo {pages} {682} (\bibinfo {year}
  {1979})}\BibitemShut {NoStop}%
\bibitem [{\citenamefont {Easther}\ \emph {et~al.}(2007)\citenamefont
  {Easther}, \citenamefont {Giblin~Jr},\ and\ \citenamefont
  {Lim}}]{eastherGravitationalWaveProduction2007}%
  \BibitemOpen
  \bibfield  {author} {\bibinfo {author} {\bibfnamefont {R.}~\bibnamefont
  {Easther}}, \bibinfo {author} {\bibfnamefont {J.~T.}\ \bibnamefont
  {Giblin~Jr}},\ and\ \bibinfo {author} {\bibfnamefont {E.~A.}\ \bibnamefont
  {Lim}},\ }\href@noop {} {\bibfield  {journal} {\bibinfo  {journal} {Physical
  Review Letters}\ }\textbf {\bibinfo {volume} {99}},\ \bibinfo {pages}
  {221301} (\bibinfo {year} {2007})}\BibitemShut {NoStop}%
\bibitem [{\citenamefont {Collaboration}(2023)}]{agazieNANOGrav15Yr2023}%
  \BibitemOpen
  \bibfield  {author} {\bibinfo {author} {\bibfnamefont {T.~N.}\ \bibnamefont
  {Collaboration}},\ }\href {https://doi.org/10.3847/2041-8213/acdac6}
  {\bibfield  {journal} {\bibinfo  {journal} {The Astrophysical Journal
  Letters}\ }\textbf {\bibinfo {volume} {951}},\ \bibinfo {pages} {L8}
  (\bibinfo {year} {2023})}\BibitemShut {NoStop}%
\bibitem [{\citenamefont {Drew}\ \emph {et~al.}(2024)\citenamefont {Drew},
  \citenamefont {Kinowski},\ and\ \citenamefont
  {Shellard}}]{drewAxionStringSource2024}%
  \BibitemOpen
  \bibfield  {author} {\bibinfo {author} {\bibfnamefont {A.}~\bibnamefont
  {Drew}}, \bibinfo {author} {\bibfnamefont {T.}~\bibnamefont {Kinowski}},\
  and\ \bibinfo {author} {\bibfnamefont {E.~P.~S.}\ \bibnamefont {Shellard}},\
  }\href {https://doi.org/10.1103/PhysRevD.110.043513} {\bibfield  {journal}
  {\bibinfo  {journal} {Physical Review D}\ }\textbf {\bibinfo {volume}
  {110}},\ \bibinfo {pages} {043513} (\bibinfo {year} {2024})}\BibitemShut
  {NoStop}%
\bibitem [{\citenamefont {Tada}\ and\ \citenamefont
  {Yamada}(2024)}]{tadaMultifieldStochasticDynamics2024}%
  \BibitemOpen
  \bibfield  {author} {\bibinfo {author} {\bibfnamefont {Y.}~\bibnamefont
  {Tada}}\ and\ \bibinfo {author} {\bibfnamefont {M.}~\bibnamefont {Yamada}},\
  }\href {https://doi.org/10.1016/j.physletb.2024.138854} {\bibfield  {journal}
  {\bibinfo  {journal} {Physics Letters B}\ }\textbf {\bibinfo {volume}
  {855}},\ \bibinfo {pages} {138854} (\bibinfo {year} {2024})}\BibitemShut
  {NoStop}%
\bibitem [{\citenamefont {Choudhury}\ \emph {et~al.}(2024)\citenamefont
  {Choudhury}, \citenamefont {Dey}, \citenamefont {Karde}, \citenamefont
  {Panda},\ and\ \citenamefont
  {Sami}}]{choudhuryPrimordialNonGaussianitySaviour2024}%
  \BibitemOpen
  \bibfield  {author} {\bibinfo {author} {\bibfnamefont {S.}~\bibnamefont
  {Choudhury}}, \bibinfo {author} {\bibfnamefont {K.}~\bibnamefont {Dey}},
  \bibinfo {author} {\bibfnamefont {A.}~\bibnamefont {Karde}}, \bibinfo
  {author} {\bibfnamefont {S.}~\bibnamefont {Panda}},\ and\ \bibinfo {author}
  {\bibfnamefont {M.}~\bibnamefont {Sami}},\ }\href
  {https://doi.org/10.1016/j.physletb.2024.138925} {\bibfield  {journal}
  {\bibinfo  {journal} {Physics Letters B}\ }\textbf {\bibinfo {volume}
  {856}},\ \bibinfo {pages} {138925} (\bibinfo {year} {2024})}\BibitemShut
  {NoStop}%
\bibitem [{\citenamefont {Launay}\ \emph
  {et~al.}(2024{\natexlab{a}})\citenamefont {Launay}, \citenamefont
  {Rigopoulos},\ and\ \citenamefont
  {Shellard}}]{launayStochasticInflationGeneral2024a}%
  \BibitemOpen
  \bibfield  {author} {\bibinfo {author} {\bibfnamefont {Y.~L.}\ \bibnamefont
  {Launay}}, \bibinfo {author} {\bibfnamefont {G.~I.}\ \bibnamefont
  {Rigopoulos}},\ and\ \bibinfo {author} {\bibfnamefont {E.~P.~S.}\
  \bibnamefont {Shellard}},\ }\href
  {https://doi.org/10.1103/PhysRevD.109.123523} {\bibfield  {journal} {\bibinfo
   {journal} {Physical Review D}\ }\textbf {\bibinfo {volume} {109}},\ \bibinfo
  {pages} {123523} (\bibinfo {year} {2024}{\natexlab{a}})}\BibitemShut
  {NoStop}%
\bibitem [{\citenamefont {Baumann}(2022)}]{baumannCosmology2022}%
  \BibitemOpen
  \bibfield  {author} {\bibinfo {author} {\bibfnamefont {D.}~\bibnamefont
  {Baumann}},\ }\href {https://doi.org/10.1017/9781108937092} {\bibinfo {title}
  {Cosmology}} (\bibinfo {year} {2022})\BibitemShut {NoStop}%
\bibitem [{\citenamefont {Paoletti}\ \emph {et~al.}(2022)\citenamefont
  {Paoletti}, \citenamefont {Finelli}, \citenamefont {Valiviita},\ and\
  \citenamefont {Hazumi}}]{paolettiPlanckBICEPKeck2022}%
  \BibitemOpen
  \bibfield  {author} {\bibinfo {author} {\bibfnamefont {D.}~\bibnamefont
  {Paoletti}}, \bibinfo {author} {\bibfnamefont {F.}~\bibnamefont {Finelli}},
  \bibinfo {author} {\bibfnamefont {J.}~\bibnamefont {Valiviita}},\ and\
  \bibinfo {author} {\bibfnamefont {M.}~\bibnamefont {Hazumi}},\ }\href
  {https://doi.org/10.1103/PhysRevD.106.083528} {\bibfield  {journal} {\bibinfo
   {journal} {Physical Review D}\ }\textbf {\bibinfo {volume} {106}},\ \bibinfo
  {pages} {083528} (\bibinfo {year} {2022})}\BibitemShut {NoStop}%
\bibitem [{\citenamefont {{BICEP/Keck
  Collaboration}}(2022)}]{bicep/keckcollaborationLatestConstraintsInflationary2022}%
  \BibitemOpen
  \bibfield  {author} {\bibinfo {author} {\bibnamefont {{BICEP/Keck
  Collaboration}}},\ }\href {https://doi.org/10.48550/arXiv.2203.16556}
  {\bibinfo {title} {The {{Latest Constraints}} on {{Inflationary B-modes}}
  from the {{BICEP}}/{{Keck Telescopes}}}} (\bibinfo {year} {2022})\BibitemShut
  {NoStop}%
\bibitem [{\citenamefont
  {Collaboration}(2022)}]{abazajianCMBS4ForecastingConstraints2022a}%
  \BibitemOpen
  \bibfield  {author} {\bibinfo {author} {\bibfnamefont {T.~C.-S.}\
  \bibnamefont {Collaboration}},\ }\href
  {https://doi.org/10.3847/1538-4357/ac1596} {\bibfield  {journal} {\bibinfo
  {journal} {The Astrophysical Journal}\ }\textbf {\bibinfo {volume} {926}},\
  \bibinfo {pages} {54} (\bibinfo {year} {2022})}\BibitemShut {NoStop}%
\bibitem [{\citenamefont {{LiteBIRD
  Collaboration}}(2023)}]{litebirdcollaborationProbingCosmicInflation2023}%
  \BibitemOpen
  \bibfield  {author} {\bibinfo {author} {\bibnamefont {{LiteBIRD
  Collaboration}}},\ }\href {https://doi.org/10.1093/ptep/ptac150} {\bibfield
  {journal} {\bibinfo  {journal} {Progress of Theoretical and Experimental
  Physics}\ }\textbf {\bibinfo {volume} {2023}},\ \bibinfo {pages} {042F01}
  (\bibinfo {year} {2023})}\BibitemShut {NoStop}%
\bibitem [{\citenamefont {Bernardeau}\ and\ \citenamefont
  {Uzan}(2002)}]{bernardeauNonGaussianityMultifieldInflation2002}%
  \BibitemOpen
  \bibfield  {author} {\bibinfo {author} {\bibfnamefont {F.}~\bibnamefont
  {Bernardeau}}\ and\ \bibinfo {author} {\bibfnamefont {J.-P.}\ \bibnamefont
  {Uzan}},\ }\href {https://doi.org/10.1103/PhysRevD.66.103506} {\bibfield
  {journal} {\bibinfo  {journal} {Physical Review D}\ }\textbf {\bibinfo
  {volume} {66}},\ \bibinfo {pages} {103506} (\bibinfo {year}
  {2002})}\BibitemShut {NoStop}%
\bibitem [{\citenamefont {Chen}\ and\ \citenamefont
  {Wang}(2010)}]{chenLargeNonGaussianitiesIntermediate2010}%
  \BibitemOpen
  \bibfield  {author} {\bibinfo {author} {\bibfnamefont {X.}~\bibnamefont
  {Chen}}\ and\ \bibinfo {author} {\bibfnamefont {Y.}~\bibnamefont {Wang}},\
  }\href {https://doi.org/10.1103/PhysRevD.81.063511} {\bibfield  {journal}
  {\bibinfo  {journal} {Physical Review D}\ }\textbf {\bibinfo {volume} {81}},\
  \bibinfo {pages} {063511} (\bibinfo {year} {2010})}\BibitemShut {NoStop}%
\bibitem [{\citenamefont {Dufaux}\ \emph {et~al.}(2007)\citenamefont {Dufaux},
  \citenamefont {Bergman}, \citenamefont {Felder}, \citenamefont {Kofman},\
  and\ \citenamefont {Uzan}}]{dufauxTheoryNumericsGravitational2007a}%
  \BibitemOpen
  \bibfield  {author} {\bibinfo {author} {\bibfnamefont {J.-F.}\ \bibnamefont
  {Dufaux}}, \bibinfo {author} {\bibfnamefont {A.}~\bibnamefont {Bergman}},
  \bibinfo {author} {\bibfnamefont {G.}~\bibnamefont {Felder}}, \bibinfo
  {author} {\bibfnamefont {L.}~\bibnamefont {Kofman}},\ and\ \bibinfo {author}
  {\bibfnamefont {J.-P.}\ \bibnamefont {Uzan}},\ }\href
  {https://doi.org/10.1103/PhysRevD.76.123517} {\bibfield  {journal} {\bibinfo
  {journal} {Physical Review D}\ }\textbf {\bibinfo {volume} {76}},\ \bibinfo
  {pages} {123517} (\bibinfo {year} {2007})}\BibitemShut {NoStop}%
\bibitem [{\citenamefont {Dufaux}\ \emph {et~al.}(2009)\citenamefont {Dufaux},
  \citenamefont {Felder}, \citenamefont {Kofman},\ and\ \citenamefont
  {Navros}}]{dufauxGravityWavesTachyonic2009}%
  \BibitemOpen
  \bibfield  {author} {\bibinfo {author} {\bibfnamefont {J.-F.}\ \bibnamefont
  {Dufaux}}, \bibinfo {author} {\bibfnamefont {G.}~\bibnamefont {Felder}},
  \bibinfo {author} {\bibfnamefont {L.}~\bibnamefont {Kofman}},\ and\ \bibinfo
  {author} {\bibfnamefont {O.}~\bibnamefont {Navros}},\ }\href
  {https://doi.org/10.1088/1475-7516/2009/03/001} {\bibfield  {journal}
  {\bibinfo  {journal} {Journal of Cosmology and Astroparticle Physics}\
  }\textbf {\bibinfo {volume} {2009}}\bibinfo  {number} { (03)},\ \bibinfo
  {pages} {001}}\BibitemShut {NoStop}%
\bibitem [{\citenamefont {Price}\ and\ \citenamefont
  {Siemens}(2008)}]{priceStochasticBackgroundsGravitational2008}%
  \BibitemOpen
\bibfield  {number} {  }\bibfield  {author} {\bibinfo {author} {\bibfnamefont
  {L.~R.}\ \bibnamefont {Price}}\ and\ \bibinfo {author} {\bibfnamefont
  {X.}~\bibnamefont {Siemens}},\ }\href
  {https://doi.org/10.1103/PhysRevD.78.063541} {\bibfield  {journal} {\bibinfo
  {journal} {Physical Review D}\ }\textbf {\bibinfo {volume} {78}},\ \bibinfo
  {pages} {063541} (\bibinfo {year} {2008})}\BibitemShut {NoStop}%
\bibitem [{\citenamefont {{Garc{\'i}a-Bellido}}\ and\ \citenamefont
  {Figueroa}(2007)}]{garcia-bellidoStochasticBackgroundGravitational2007a}%
  \BibitemOpen
  \bibfield  {author} {\bibinfo {author} {\bibfnamefont {J.}~\bibnamefont
  {{Garc{\'i}a-Bellido}}}\ and\ \bibinfo {author} {\bibfnamefont {D.~G.}\
  \bibnamefont {Figueroa}},\ }\href
  {https://doi.org/10.1103/PhysRevLett.98.061302} {\bibfield  {journal}
  {\bibinfo  {journal} {Physical Review Letters}\ }\textbf {\bibinfo {volume}
  {98}},\ \bibinfo {pages} {061302} (\bibinfo {year} {2007})}\BibitemShut
  {NoStop}%
\bibitem [{\citenamefont {Khlebnikov}\ and\ \citenamefont
  {Tkachev}(1997)}]{khlebnikovRelicGravitationalWaves1997}%
  \BibitemOpen
  \bibfield  {author} {\bibinfo {author} {\bibfnamefont {S.}~\bibnamefont
  {Khlebnikov}}\ and\ \bibinfo {author} {\bibfnamefont {I.}~\bibnamefont
  {Tkachev}},\ }\href {https://doi.org/10.1103/PhysRevD.56.653} {\bibfield
  {journal} {\bibinfo  {journal} {Physical Review D}\ }\textbf {\bibinfo
  {volume} {56}},\ \bibinfo {pages} {653} (\bibinfo {year} {1997})}\BibitemShut
  {NoStop}%
\bibitem [{\citenamefont {Iacconi}\ and\ \citenamefont
  {Mulryne}(2023)}]{iacconiMultifieldInflationLarge2023}%
  \BibitemOpen
  \bibfield  {author} {\bibinfo {author} {\bibfnamefont {L.}~\bibnamefont
  {Iacconi}}\ and\ \bibinfo {author} {\bibfnamefont {D.~J.}\ \bibnamefont
  {Mulryne}},\ }\href {https://doi.org/10.1088/1475-7516/2023/09/033}
  {\bibfield  {journal} {\bibinfo  {journal} {Journal of Cosmology and
  Astroparticle Physics}\ }\textbf {\bibinfo {volume} {2023}}\bibinfo  {number}
  { (09)},\ \bibinfo {pages} {033}}\BibitemShut {NoStop}%
\bibitem [{\citenamefont {Felder}\ and\ \citenamefont
  {Tkachev}(2008)}]{felderLATTICEEASYProgramLattice2008}%
  \BibitemOpen
\bibfield  {number} {  }\bibfield  {author} {\bibinfo {author} {\bibfnamefont
  {G.}~\bibnamefont {Felder}}\ and\ \bibinfo {author} {\bibfnamefont
  {I.}~\bibnamefont {Tkachev}},\ }\href
  {https://doi.org/10.1016/j.cpc.2008.02.009} {\bibfield  {journal} {\bibinfo
  {journal} {Computer Physics Communications}\ }\textbf {\bibinfo {volume}
  {178}},\ \bibinfo {pages} {929} (\bibinfo {year} {2008})}\BibitemShut
  {NoStop}%
\bibitem [{\citenamefont {Jr}\ \emph {et~al.}(2010)\citenamefont {Jr},
  \citenamefont {Price},\ and\ \citenamefont
  {Siemens}}]{giblinGravitationalRadiationPreheating2010}%
  \BibitemOpen
  \bibfield  {author} {\bibinfo {author} {\bibfnamefont {J.~T.~G.}\
  \bibnamefont {Jr}}, \bibinfo {author} {\bibfnamefont {L.~R.}\ \bibnamefont
  {Price}},\ and\ \bibinfo {author} {\bibfnamefont {X.}~\bibnamefont
  {Siemens}},\ }\href {https://doi.org/10.1088/1475-7516/2010/08/012}
  {\bibfield  {journal} {\bibinfo  {journal} {Journal of Cosmology and
  Astroparticle Physics}\ }\textbf {\bibinfo {volume} {2010}}\bibinfo  {number}
  { (08)},\ \bibinfo {pages} {012}}\BibitemShut {NoStop}%
\bibitem [{\citenamefont {Huang}(2011)}]{huangArtLatticeGravity2011}%
  \BibitemOpen
\bibfield  {number} {  }\bibfield  {author} {\bibinfo {author} {\bibfnamefont
  {Z.}~\bibnamefont {Huang}},\ }\href
  {https://doi.org/10.1103/PhysRevD.83.123509} {\bibfield  {journal} {\bibinfo
  {journal} {Physical Review D}\ }\textbf {\bibinfo {volume} {83}},\ \bibinfo
  {pages} {123509} (\bibinfo {year} {2011})}\BibitemShut {NoStop}%
\bibitem [{\citenamefont {Clough}\ \emph {et~al.}(2015)\citenamefont {Clough},
  \citenamefont {Figueras}, \citenamefont {Finkel}, \citenamefont {Kunesch},
  \citenamefont {Lim},\ and\ \citenamefont
  {Tunyasuvunakool}}]{cloughGRChomboNumericalRelativity2015a}%
  \BibitemOpen
  \bibfield  {author} {\bibinfo {author} {\bibfnamefont {K.}~\bibnamefont
  {Clough}}, \bibinfo {author} {\bibfnamefont {P.}~\bibnamefont {Figueras}},
  \bibinfo {author} {\bibfnamefont {H.}~\bibnamefont {Finkel}}, \bibinfo
  {author} {\bibfnamefont {M.}~\bibnamefont {Kunesch}}, \bibinfo {author}
  {\bibfnamefont {E.~A.}\ \bibnamefont {Lim}},\ and\ \bibinfo {author}
  {\bibfnamefont {S.}~\bibnamefont {Tunyasuvunakool}},\ }\href
  {https://doi.org/10.1088/0264-9381/32/24/245011} {\bibfield  {journal}
  {\bibinfo  {journal} {Class. Quant. Grav.}\ }\textbf {\bibinfo {volume}
  {32}},\ \bibinfo {pages} {245011} (\bibinfo {year} {2015})}\BibitemShut
  {NoStop}%
\bibitem [{\citenamefont {Figueroa}\ \emph {et~al.}(2023)\citenamefont
  {Figueroa}, \citenamefont {Florio}, \citenamefont {Torrenti},\ and\
  \citenamefont {Valkenburg}}]{figueroaCosmoLatticeModernCode2023}%
  \BibitemOpen
  \bibfield  {author} {\bibinfo {author} {\bibfnamefont {D.~G.}\ \bibnamefont
  {Figueroa}}, \bibinfo {author} {\bibfnamefont {A.}~\bibnamefont {Florio}},
  \bibinfo {author} {\bibfnamefont {F.}~\bibnamefont {Torrenti}},\ and\
  \bibinfo {author} {\bibfnamefont {W.}~\bibnamefont {Valkenburg}},\ }\href
  {https://doi.org/10.1016/j.cpc.2022.108586} {\bibfield  {journal} {\bibinfo
  {journal} {Computer Physics Communications}\ }\textbf {\bibinfo {volume}
  {283}},\ \bibinfo {pages} {108586} (\bibinfo {year} {2023})}\BibitemShut
  {NoStop}%
\bibitem [{\citenamefont {{van de Vis}}\ \emph {et~al.}(2020)\citenamefont
  {{van de Vis}}, \citenamefont {Nguyen}, \citenamefont {Sfakianakis},
  \citenamefont {Giblin},\ and\ \citenamefont
  {Kaiser}}]{vandevisTimeScalesNonlinear2020}%
  \BibitemOpen
  \bibfield  {author} {\bibinfo {author} {\bibfnamefont {J.}~\bibnamefont {{van
  de Vis}}}, \bibinfo {author} {\bibfnamefont {R.}~\bibnamefont {Nguyen}},
  \bibinfo {author} {\bibfnamefont {E.~I.}\ \bibnamefont {Sfakianakis}},
  \bibinfo {author} {\bibfnamefont {J.~T.}\ \bibnamefont {Giblin}},\ and\
  \bibinfo {author} {\bibfnamefont {D.~I.}\ \bibnamefont {Kaiser}},\ }\href
  {https://doi.org/10.1103/PhysRevD.102.043528} {\bibfield  {journal} {\bibinfo
   {journal} {Physical Review D}\ }\textbf {\bibinfo {volume} {102}},\ \bibinfo
  {pages} {043528} (\bibinfo {year} {2020})}\BibitemShut {NoStop}%
\bibitem [{\citenamefont {Adshead}\ \emph
  {et~al.}(2024{\natexlab{a}})\citenamefont {Adshead}, \citenamefont {Giblin},\
  and\ \citenamefont {Tishue}}]{adsheadGravitationalWavesKinetic2024a}%
  \BibitemOpen
  \bibfield  {author} {\bibinfo {author} {\bibfnamefont {P.}~\bibnamefont
  {Adshead}}, \bibinfo {author} {\bibfnamefont {J.~T.}\ \bibnamefont
  {Giblin}},\ and\ \bibinfo {author} {\bibfnamefont {A.}~\bibnamefont
  {Tishue}},\ }\href {https://doi.org/10.1103/PhysRevD.110.043536} {\bibfield
  {journal} {\bibinfo  {journal} {Physical Review D}\ }\textbf {\bibinfo
  {volume} {110}},\ \bibinfo {pages} {043536} (\bibinfo {year}
  {2024}{\natexlab{a}})}\BibitemShut {NoStop}%
\bibitem [{\citenamefont {Figueroa}\ and\ \citenamefont
  {Loayza}(2024)}]{figueroaGeometricReheatingUniverse2024}%
  \BibitemOpen
  \bibfield  {author} {\bibinfo {author} {\bibfnamefont {D.~G.}\ \bibnamefont
  {Figueroa}}\ and\ \bibinfo {author} {\bibfnamefont {N.}~\bibnamefont
  {Loayza}},\ }\href {https://doi.org/10.48550/arXiv.2406.02689} {\bibinfo
  {title} {Geometric reheating of the {{Universe}}}} (\bibinfo {year} {2024}),\
  \Eprint {https://arxiv.org/abs/2406.02689} {arXiv:2406.02689 [astro-ph]}
  \BibitemShut {NoStop}%
\bibitem [{\citenamefont {Dahl}\ \emph {et~al.}(2022)\citenamefont {Dahl},
  \citenamefont {Hindmarsh}, \citenamefont {Rummukainen},\ and\ \citenamefont
  {Weir}}]{dahlDecayAcousticTurbulence2022}%
  \BibitemOpen
  \bibfield  {author} {\bibinfo {author} {\bibfnamefont {J.}~\bibnamefont
  {Dahl}}, \bibinfo {author} {\bibfnamefont {M.}~\bibnamefont {Hindmarsh}},
  \bibinfo {author} {\bibfnamefont {K.}~\bibnamefont {Rummukainen}},\ and\
  \bibinfo {author} {\bibfnamefont {D.~J.}\ \bibnamefont {Weir}},\ }\href
  {https://doi.org/10.1103/PhysRevD.106.063511} {\bibfield  {journal} {\bibinfo
   {journal} {Physical Review D}\ }\textbf {\bibinfo {volume} {106}},\ \bibinfo
  {pages} {063511} (\bibinfo {year} {2022})}\BibitemShut {NoStop}%
\bibitem [{\citenamefont {Figueroa}\ \emph {et~al.}(2024)\citenamefont
  {Figueroa}, \citenamefont {Lizarraga}, \citenamefont {Loayza}, \citenamefont
  {Urio},\ and\ \citenamefont
  {Urrestilla}}]{figueroaNonlinearDynamicsAxion2024}%
  \BibitemOpen
  \bibfield  {author} {\bibinfo {author} {\bibfnamefont {D.~G.}\ \bibnamefont
  {Figueroa}}, \bibinfo {author} {\bibfnamefont {J.}~\bibnamefont {Lizarraga}},
  \bibinfo {author} {\bibfnamefont {N.}~\bibnamefont {Loayza}}, \bibinfo
  {author} {\bibfnamefont {A.}~\bibnamefont {Urio}},\ and\ \bibinfo {author}
  {\bibfnamefont {J.}~\bibnamefont {Urrestilla}},\ }\href
  {https://doi.org/10.48550/arXiv.2411.16368} {\bibinfo {title} {The non-linear
  dynamics of axion inflation: A detailed lattice study}} (\bibinfo {year}
  {2024}),\ \Eprint {https://arxiv.org/abs/2411.16368} {arXiv:2411.16368
  [astro-ph]} \BibitemShut {NoStop}%
\bibitem [{\citenamefont {Nguyen}\ \emph {et~al.}(2019)\citenamefont {Nguyen},
  \citenamefont {{van de Vis}}, \citenamefont {Sfakianakis}, \citenamefont
  {Giblin},\ and\ \citenamefont
  {Kaiser}}]{nguyenNonlinearDynamicsPreheating2019}%
  \BibitemOpen
  \bibfield  {author} {\bibinfo {author} {\bibfnamefont {R.}~\bibnamefont
  {Nguyen}}, \bibinfo {author} {\bibfnamefont {J.}~\bibnamefont {{van de
  Vis}}}, \bibinfo {author} {\bibfnamefont {E.~I.}\ \bibnamefont
  {Sfakianakis}}, \bibinfo {author} {\bibfnamefont {J.~T.}\ \bibnamefont
  {Giblin}},\ and\ \bibinfo {author} {\bibfnamefont {D.~I.}\ \bibnamefont
  {Kaiser}},\ }\href {https://doi.org/10.1103/PhysRevLett.123.171301}
  {\bibfield  {journal} {\bibinfo  {journal} {Physical Review Letters}\
  }\textbf {\bibinfo {volume} {123}},\ \bibinfo {pages} {171301} (\bibinfo
  {year} {2019})}\BibitemShut {NoStop}%
\bibitem [{\citenamefont {{Bastero-Gil}}\ \emph {et~al.}(2008)\citenamefont
  {{Bastero-Gil}}, \citenamefont {Tristram}, \citenamefont
  {{Macias-P{\'e}rez}},\ and\ \citenamefont
  {Santos}}]{bastero-gilNonlinearPreheatingScalar2008}%
  \BibitemOpen
  \bibfield  {author} {\bibinfo {author} {\bibfnamefont {M.}~\bibnamefont
  {{Bastero-Gil}}}, \bibinfo {author} {\bibfnamefont {M.}~\bibnamefont
  {Tristram}}, \bibinfo {author} {\bibfnamefont {J.}~\bibnamefont
  {{Macias-P{\'e}rez}}},\ and\ \bibinfo {author} {\bibfnamefont
  {D.}~\bibnamefont {Santos}},\ }\href
  {https://doi.org/10.1103/PhysRevD.77.023520} {\bibfield  {journal} {\bibinfo
  {journal} {Physical Review D}\ }\textbf {\bibinfo {volume} {77}},\ \bibinfo
  {pages} {023520} (\bibinfo {year} {2008})}\BibitemShut {NoStop}%
\bibitem [{\citenamefont {Giblin~Jr}\ and\ \citenamefont
  {Tishue}(2019)}]{giblinjrPreheatingFullGeneral2019}%
  \BibitemOpen
  \bibfield  {author} {\bibinfo {author} {\bibfnamefont {J.~T.}\ \bibnamefont
  {Giblin~Jr}}\ and\ \bibinfo {author} {\bibfnamefont {A.~J.}\ \bibnamefont
  {Tishue}},\ }\href {https://doi.org/10.1103/PhysRevD.100.063543} {\bibfield
  {journal} {\bibinfo  {journal} {Physical Review D}\ }\textbf {\bibinfo
  {volume} {100}},\ \bibinfo {pages} {063543} (\bibinfo {year} {2019})},\
  \Eprint {https://arxiv.org/abs/1907.10601} {arXiv:1907.10601 [astro-ph,
  physics:gr-qc]} \BibitemShut {NoStop}%
\bibitem [{\citenamefont
  {Joana}(2022{\natexlab{a}})}]{joanaCosmicInhomogeneitiesEarly2022}%
  \BibitemOpen
  \bibfield  {author} {\bibinfo {author} {\bibfnamefont {C.}~\bibnamefont
  {Joana}},\ }\href@noop {} {\bibinfo {title} {Cosmic inhomogeneities in the
  early {{Universe}}: {{A}} numerical relativity approach}} (\bibinfo {year}
  {2022}{\natexlab{a}}),\ \Eprint {https://arxiv.org/abs/2211.03534}
  {arXiv:2211.03534 [astro-ph]} \BibitemShut {NoStop}%
\bibitem [{\citenamefont {Adshead}\ \emph
  {et~al.}(2024{\natexlab{b}})\citenamefont {Adshead}, \citenamefont {Giblin},
  \citenamefont {Grutkoski},\ and\ \citenamefont
  {Weiner}}]{adsheadGaugePreheatingFull2024}%
  \BibitemOpen
  \bibfield  {author} {\bibinfo {author} {\bibfnamefont {P.}~\bibnamefont
  {Adshead}}, \bibinfo {author} {\bibfnamefont {J.~T.}\ \bibnamefont {Giblin}},
  \bibinfo {author} {\bibfnamefont {R.}~\bibnamefont {Grutkoski}},\ and\
  \bibinfo {author} {\bibfnamefont {Z.~J.}\ \bibnamefont {Weiner}},\ }\href
  {https://doi.org/10.1088/1475-7516/2024/03/017} {\bibfield  {journal}
  {\bibinfo  {journal} {Journal of Cosmology and Astroparticle Physics}\
  }\textbf {\bibinfo {volume} {2024}}\bibinfo  {number} { (03)},\ \bibinfo
  {pages} {017}}\BibitemShut {NoStop}%
\bibitem [{\citenamefont {Raveendran}\ and\ \citenamefont
  {Sriramkumar}(2017)}]{raveendranNumericalEvaluationTensor2017}%
  \BibitemOpen
\bibfield  {number} {  }\bibfield  {author} {\bibinfo {author} {\bibfnamefont
  {R.~N.}\ \bibnamefont {Raveendran}}\ and\ \bibinfo {author} {\bibfnamefont
  {L.}~\bibnamefont {Sriramkumar}},\ }\href
  {https://doi.org/10.1088/1475-7516/2017/07/035} {\bibfield  {journal}
  {\bibinfo  {journal} {Journal of Cosmology and Astroparticle Physics}\
  }\textbf {\bibinfo {volume} {2017}}\bibfield  {number} {\bibinfo  {number} {
  (07)},\ \bibinfo {pages} {035}},\ }\Eprint {https://arxiv.org/abs/1611.00473}
  {arXiv:1611.00473 [astro-ph, physics:gr-qc, physics:hep-th]} \BibitemShut
  {NoStop}%
\bibitem [{\citenamefont {Bari}\ \emph {et~al.}(2024)\citenamefont {Bari},
  \citenamefont {Bartolo}, \citenamefont {Dom{\`e}nech},\ and\ \citenamefont
  {Matarrese}}]{bariGravitationalWavesInduced2024}%
  \BibitemOpen
  \bibfield  {author} {\bibinfo {author} {\bibfnamefont {P.}~\bibnamefont
  {Bari}}, \bibinfo {author} {\bibfnamefont {N.}~\bibnamefont {Bartolo}},
  \bibinfo {author} {\bibfnamefont {G.}~\bibnamefont {Dom{\`e}nech}},\ and\
  \bibinfo {author} {\bibfnamefont {S.}~\bibnamefont {Matarrese}},\ }\href
  {https://doi.org/10.1103/PhysRevD.109.023509} {\bibfield  {journal} {\bibinfo
   {journal} {Physical Review D}\ }\textbf {\bibinfo {volume} {109}},\ \bibinfo
  {pages} {023509} (\bibinfo {year} {2024})}\BibitemShut {NoStop}%
\bibitem [{\citenamefont {Clough}\ \emph {et~al.}(2018)\citenamefont {Clough},
  \citenamefont {Flauger},\ and\ \citenamefont
  {Lim}}]{cloughRobustnessInflationLarge2018}%
  \BibitemOpen
  \bibfield  {author} {\bibinfo {author} {\bibfnamefont {K.}~\bibnamefont
  {Clough}}, \bibinfo {author} {\bibfnamefont {R.}~\bibnamefont {Flauger}},\
  and\ \bibinfo {author} {\bibfnamefont {E.~A.}\ \bibnamefont {Lim}},\ }\href
  {https://doi.org/10.1088/1475-7516/2018/05/065} {\bibfield  {journal}
  {\bibinfo  {journal} {Journal of Cosmology and Astroparticle Physics}\
  }\textbf {\bibinfo {volume} {2018}}\bibfield  {number} {\bibinfo  {number} {
  (05)},\ \bibinfo {pages} {065}},\ }\Eprint {https://arxiv.org/abs/1712.07352}
  {arXiv:1712.07352 [astro-ph, physics:gr-qc, physics:hep-th]} \BibitemShut
  {NoStop}%
\bibitem [{\citenamefont {Clough}\ and\ \citenamefont
  {Niemeyer}(2018)}]{cloughDifficultyGeneratingGravitational2018}%
  \BibitemOpen
  \bibfield  {author} {\bibinfo {author} {\bibfnamefont {K.}~\bibnamefont
  {Clough}}\ and\ \bibinfo {author} {\bibfnamefont {J.~C.}\ \bibnamefont
  {Niemeyer}},\ }\href {https://doi.org/10.1088/1361-6382/aad7f0} {\bibfield
  {journal} {\bibinfo  {journal} {Classical and Quantum Gravity}\ }\textbf
  {\bibinfo {volume} {35}},\ \bibinfo {pages} {187001} (\bibinfo {year}
  {2018})},\ \Eprint {https://arxiv.org/abs/1803.10719} {arXiv:1803.10719
  [gr-qc]} \BibitemShut {NoStop}%
\bibitem [{\citenamefont {Aurrekoetxea}\ \emph {et~al.}(2020)\citenamefont
  {Aurrekoetxea}, \citenamefont {Clough}, \citenamefont {Flauger},\ and\
  \citenamefont {Lim}}]{aurrekoetxeaEffectsPotentialShape2020}%
  \BibitemOpen
  \bibfield  {author} {\bibinfo {author} {\bibfnamefont {J.~C.}\ \bibnamefont
  {Aurrekoetxea}}, \bibinfo {author} {\bibfnamefont {K.}~\bibnamefont
  {Clough}}, \bibinfo {author} {\bibfnamefont {R.}~\bibnamefont {Flauger}},\
  and\ \bibinfo {author} {\bibfnamefont {E.~A.}\ \bibnamefont {Lim}},\ }\href
  {https://doi.org/10.1088/1475-7516/2020/05/030} {\bibfield  {journal}
  {\bibinfo  {journal} {Journal of Cosmology and Astroparticle Physics}\
  }\textbf {\bibinfo {volume} {2020}}\bibfield  {number} {\bibinfo  {number} {
  (05)},\ \bibinfo {pages} {030}},\ }\Eprint {https://arxiv.org/abs/1910.12547}
  {arXiv:1910.12547 [astro-ph, physics:gr-qc, physics:hep-th]} \BibitemShut
  {NoStop}%
\bibitem [{\citenamefont
  {Joana}(2022{\natexlab{b}})}]{joanaGravitationalDynamicsHiggs2022}%
  \BibitemOpen
  \bibfield  {author} {\bibinfo {author} {\bibfnamefont {C.}~\bibnamefont
  {Joana}},\ }\href {https://doi.org/10.1103/PhysRevD.106.023504} {\bibfield
  {journal} {\bibinfo  {journal} {Physical Review D}\ }\textbf {\bibinfo
  {volume} {106}},\ \bibinfo {pages} {023504} (\bibinfo {year}
  {2022}{\natexlab{b}})},\ \Eprint {https://arxiv.org/abs/2202.07604}
  {arXiv:2202.07604 [astro-ph, physics:gr-qc, physics:hep-ph, physics:hep-th]}
  \BibitemShut {NoStop}%
\bibitem [{\citenamefont {Figueroa}\ \emph {et~al.}(2021)\citenamefont
  {Figueroa}, \citenamefont {Florio}, \citenamefont {Torrenti},\ and\
  \citenamefont {Valkenburg}}]{figueroaArtSimulatingEarly2021}%
  \BibitemOpen
  \bibfield  {author} {\bibinfo {author} {\bibfnamefont {D.~G.}\ \bibnamefont
  {Figueroa}}, \bibinfo {author} {\bibfnamefont {A.}~\bibnamefont {Florio}},
  \bibinfo {author} {\bibfnamefont {F.}~\bibnamefont {Torrenti}},\ and\
  \bibinfo {author} {\bibfnamefont {W.}~\bibnamefont {Valkenburg}},\ }\href
  {https://doi.org/10.1088/1475-7516/2021/04/035} {\bibfield  {journal}
  {\bibinfo  {journal} {Journal of Cosmology and Astroparticle Physics}\
  }\textbf {\bibinfo {volume} {2021}}\bibfield  {number} {\bibinfo  {number} {
  (04)},\ \bibinfo {pages} {035}},\ }\Eprint {https://arxiv.org/abs/2006.15122}
  {arXiv:2006.15122 [astro-ph]} \BibitemShut {NoStop}%
\bibitem [{\citenamefont {Khlebnikov}\ and\ \citenamefont
  {Tkachev}(1996)}]{khlebnikovClassicalDecayInflaton1996}%
  \BibitemOpen
  \bibfield  {author} {\bibinfo {author} {\bibfnamefont {S.~{\relax Yu}.}\
  \bibnamefont {Khlebnikov}}\ and\ \bibinfo {author} {\bibfnamefont {I.~I.}\
  \bibnamefont {Tkachev}},\ }\href {https://doi.org/10.1103/PhysRevLett.77.219}
  {\bibfield  {journal} {\bibinfo  {journal} {Physical Review Letters}\
  }\textbf {\bibinfo {volume} {77}},\ \bibinfo {pages} {219} (\bibinfo {year}
  {1996})}\BibitemShut {NoStop}%
\bibitem [{\citenamefont
  {Maldacena}(2003)}]{maldacenaNongaussianFeaturesPrimordial2003a}%
  \BibitemOpen
  \bibfield  {author} {\bibinfo {author} {\bibfnamefont {J.}~\bibnamefont
  {Maldacena}},\ }\href {https://doi.org/10.1088/1126-6708/2003/05/013}
  {\bibfield  {journal} {\bibinfo  {journal} {Journal of High Energy Physics}\
  }\textbf {\bibinfo {volume} {2003}},\ \bibinfo {pages} {013} (\bibinfo {year}
  {2003})}\BibitemShut {NoStop}%
\bibitem [{\citenamefont {Clough}(2018)}]{cloughScalarFieldsNumerical2018}%
  \BibitemOpen
  \bibfield  {author} {\bibinfo {author} {\bibfnamefont {K.}~\bibnamefont
  {Clough}},\ }\href {https://doi.org/10.1007/978-3-319-92672-8} {\emph
  {\bibinfo {title} {Scalar {{Fields}} in {{Numerical General Relativity}}}}},\
  Springer {{Theses}}\ (\bibinfo  {publisher} {Springer International
  Publishing},\ \bibinfo {address} {Cham},\ \bibinfo {year} {2018})\BibitemShut
  {NoStop}%
\bibitem [{\citenamefont {Carroll}(1997)}]{carrollLectureNotesGeneral1997}%
  \BibitemOpen
  \bibfield  {author} {\bibinfo {author} {\bibfnamefont {S.~M.}\ \bibnamefont
  {Carroll}},\ }\href {https://doi.org/10.48550/arXiv.gr-qc/9712019} {\bibinfo
  {title} {Lecture {{Notes}} on {{General Relativity}}}} (\bibinfo {year}
  {1997}),\ \Eprint {https://arxiv.org/abs/gr-qc/9712019} {arXiv:gr-qc/9712019}
  \BibitemShut {NoStop}%
\bibitem [{\citenamefont
  {Isi}(2022)}]{isiParametrizingGravitationalwavePolarizations2022}%
  \BibitemOpen
  \bibfield  {author} {\bibinfo {author} {\bibfnamefont {M.}~\bibnamefont
  {Isi}},\ }\href@noop {} {\bibinfo {title} {Parametrizing gravitational-wave
  polarizations}} (\bibinfo {year} {2022}),\ \Eprint
  {https://arxiv.org/abs/2208.03372} {arXiv:2208.03372 [astro-ph,
  physics:gr-qc]} \BibitemShut {NoStop}%
\bibitem [{\citenamefont {Aurrekoetxea}\ \emph {et~al.}(2024)\citenamefont
  {Aurrekoetxea}, \citenamefont {Clough},\ and\ \citenamefont
  {Lim}}]{aurrekoetxeaCosmologyUsingNumerical2024}%
  \BibitemOpen
  \bibfield  {author} {\bibinfo {author} {\bibfnamefont {J.~C.}\ \bibnamefont
  {Aurrekoetxea}}, \bibinfo {author} {\bibfnamefont {K.}~\bibnamefont
  {Clough}},\ and\ \bibinfo {author} {\bibfnamefont {E.~A.}\ \bibnamefont
  {Lim}},\ }\href {https://doi.org/10.48550/arXiv.2409.01939} {\bibinfo {title}
  {Cosmology using numerical relativity}} (\bibinfo {year} {2024}),\ \Eprint
  {https://arxiv.org/abs/2409.01939} {arXiv:2409.01939 [astro-ph,
  physics:gr-qc]} \BibitemShut {NoStop}%
\bibitem [{\citenamefont {Golovnev}\ and\ \citenamefont
  {Vanchurin}(2009)}]{golovnevCosmologicalPerturbationsVector2009}%
  \BibitemOpen
  \bibfield  {author} {\bibinfo {author} {\bibfnamefont {A.}~\bibnamefont
  {Golovnev}}\ and\ \bibinfo {author} {\bibfnamefont {V.}~\bibnamefont
  {Vanchurin}},\ }\href {https://doi.org/10.1103/PhysRevD.79.103524} {\bibfield
   {journal} {\bibinfo  {journal} {Physical Review D}\ }\textbf {\bibinfo
  {volume} {79}},\ \bibinfo {pages} {103524} (\bibinfo {year} {2009})},\
  \Eprint {https://arxiv.org/abs/0903.2977} {arXiv:0903.2977 [astro-ph,
  physics:gr-qc, physics:hep-th]} \BibitemShut {NoStop}%
\bibitem [{\citenamefont {Mukhanov}\ \emph {et~al.}(1992)\citenamefont
  {Mukhanov}, \citenamefont {Feldman},\ and\ \citenamefont
  {Brandenberger}}]{Mukhanov:1990me}%
  \BibitemOpen
  \bibfield  {author} {\bibinfo {author} {\bibfnamefont {V.~F.}\ \bibnamefont
  {Mukhanov}}, \bibinfo {author} {\bibfnamefont {H.~A.}\ \bibnamefont
  {Feldman}},\ and\ \bibinfo {author} {\bibfnamefont {R.~H.}\ \bibnamefont
  {Brandenberger}},\ }\href {https://doi.org/10.1016/0370-1573(92)90044-Z}
  {\bibfield  {journal} {\bibinfo  {journal} {Phys. Rept.}\ }\textbf {\bibinfo
  {volume} {215}},\ \bibinfo {pages} {203} (\bibinfo {year}
  {1992})}\BibitemShut {NoStop}%
\bibitem [{\citenamefont {Kodama}\ and\ \citenamefont
  {Sasaki}(1984)}]{Kodama:1984ziu}%
  \BibitemOpen
  \bibfield  {author} {\bibinfo {author} {\bibfnamefont {H.}~\bibnamefont
  {Kodama}}\ and\ \bibinfo {author} {\bibfnamefont {M.}~\bibnamefont
  {Sasaki}},\ }\href {https://doi.org/10.1143/PTPS.78.1} {\bibfield  {journal}
  {\bibinfo  {journal} {Prog. Theor. Phys. Suppl.}\ }\textbf {\bibinfo {volume}
  {78}},\ \bibinfo {pages} {1} (\bibinfo {year} {1984})}\BibitemShut {NoStop}%
\bibitem [{\citenamefont {Baumgarte}\ and\ \citenamefont
  {Shapiro}(1998)}]{baumgarteNumericalIntegrationEinsteins1998}%
  \BibitemOpen
  \bibfield  {author} {\bibinfo {author} {\bibfnamefont {T.~W.}\ \bibnamefont
  {Baumgarte}}\ and\ \bibinfo {author} {\bibfnamefont {S.~L.}\ \bibnamefont
  {Shapiro}},\ }\href {https://doi.org/10.48550/arXiv.gr-qc/9810065} {\bibinfo
  {title} {On the {{Numerical Integration}} of {{Einstein}}'s {{Field
  Equations}}}} (\bibinfo {year} {1998}),\ \Eprint
  {https://arxiv.org/abs/gr-qc/9810065} {arXiv:gr-qc/9810065} \BibitemShut
  {NoStop}%
\bibitem [{\citenamefont
  {Shibata}(1995)}]{shibataEvolutionThreedimensionalGravitational1995a}%
  \BibitemOpen
  \bibfield  {author} {\bibinfo {author} {\bibfnamefont {M.}~\bibnamefont
  {Shibata}},\ }\href {https://doi.org/10.1103/PhysRevD.52.5428} {\bibfield
  {journal} {\bibinfo  {journal} {Physical Review D}\ }\textbf {\bibinfo
  {volume} {52}},\ \bibinfo {pages} {5428} (\bibinfo {year}
  {1995})}\BibitemShut {NoStop}%
\bibitem [{\citenamefont {Arnowitt}\ \emph {et~al.}(2004)\citenamefont
  {Arnowitt}, \citenamefont {Deser},\ and\ \citenamefont
  {Misner}}]{arnowittDynamicsGeneralRelativity2004}%
  \BibitemOpen
  \bibfield  {author} {\bibinfo {author} {\bibfnamefont {R.}~\bibnamefont
  {Arnowitt}}, \bibinfo {author} {\bibfnamefont {S.}~\bibnamefont {Deser}},\
  and\ \bibinfo {author} {\bibfnamefont {C.~W.}\ \bibnamefont {Misner}},\
  }\href {https://doi.org/10.48550/arXiv.gr-qc/0405109} {\bibinfo {title} {The
  {{Dynamics}} of {{General Relativity}}}} (\bibinfo {year} {2004}),\ \Eprint
  {https://arxiv.org/abs/gr-qc/0405109} {arXiv:gr-qc/0405109} \BibitemShut
  {NoStop}%
\bibitem [{\citenamefont {Baumgarte}\ and\ \citenamefont
  {Shapiro}(2010)}]{baumgarteNumericalRelativitySolving2010}%
  \BibitemOpen
  \bibfield  {author} {\bibinfo {author} {\bibfnamefont {T.~W.}\ \bibnamefont
  {Baumgarte}}\ and\ \bibinfo {author} {\bibfnamefont {S.~L.}\ \bibnamefont
  {Shapiro}},\ }\href {https://doi.org/10.1017/CBO9781139193344} {\emph
  {\bibinfo {title} {Numerical {{Relativity}}: {{Solving Einstein}}'s
  {{Equations}} on the {{Computer}}}}}\ (\bibinfo  {publisher} {Cambridge
  University Press},\ \bibinfo {address} {Cambridge},\ \bibinfo {year}
  {2010})\BibitemShut {NoStop}%
\bibitem [{\citenamefont {Radia}\ \emph {et~al.}(2022)\citenamefont {Radia},
  \citenamefont {Sperhake}, \citenamefont {Drew}, \citenamefont {Clough},
  \citenamefont {Figueras}, \citenamefont {Lim}, \citenamefont {Ripley},
  \citenamefont {Aurrekoetxea}, \citenamefont {Fran{\c c}a},\ and\
  \citenamefont {Helfer}}]{radiaLessonsAdaptiveMesh2022}%
  \BibitemOpen
  \bibfield  {author} {\bibinfo {author} {\bibfnamefont {M.}~\bibnamefont
  {Radia}}, \bibinfo {author} {\bibfnamefont {U.}~\bibnamefont {Sperhake}},
  \bibinfo {author} {\bibfnamefont {A.}~\bibnamefont {Drew}}, \bibinfo {author}
  {\bibfnamefont {K.}~\bibnamefont {Clough}}, \bibinfo {author} {\bibfnamefont
  {P.}~\bibnamefont {Figueras}}, \bibinfo {author} {\bibfnamefont {E.~A.}\
  \bibnamefont {Lim}}, \bibinfo {author} {\bibfnamefont {J.~L.}\ \bibnamefont
  {Ripley}}, \bibinfo {author} {\bibfnamefont {J.~C.}\ \bibnamefont
  {Aurrekoetxea}}, \bibinfo {author} {\bibfnamefont {T.}~\bibnamefont {Fran{\c
  c}a}},\ and\ \bibinfo {author} {\bibfnamefont {T.}~\bibnamefont {Helfer}},\
  }\href {https://doi.org/10.1088/1361-6382/ac6fa9} {\bibfield  {journal}
  {\bibinfo  {journal} {Classical and Quantum Gravity}\ }\textbf {\bibinfo
  {volume} {39}},\ \bibinfo {pages} {135006} (\bibinfo {year} {2022})},\
  \Eprint {https://arxiv.org/abs/2112.10567} {arXiv:2112.10567 [astro-ph,
  physics:gr-qc, physics:hep-ph]} \BibitemShut {NoStop}%
\bibitem [{\citenamefont {Bona}\ \emph {et~al.}(1995)\citenamefont {Bona},
  \citenamefont {Mass{\'o}}, \citenamefont {Seidel},\ and\ \citenamefont
  {Stela}}]{bonaNewFormalismNumerical1995}%
  \BibitemOpen
  \bibfield  {author} {\bibinfo {author} {\bibfnamefont {C.}~\bibnamefont
  {Bona}}, \bibinfo {author} {\bibfnamefont {J.}~\bibnamefont {Mass{\'o}}},
  \bibinfo {author} {\bibfnamefont {E.}~\bibnamefont {Seidel}},\ and\ \bibinfo
  {author} {\bibfnamefont {J.}~\bibnamefont {Stela}},\ }\href
  {https://doi.org/10.1103/PhysRevLett.75.600} {\bibfield  {journal} {\bibinfo
  {journal} {Physical Review Letters}\ }\textbf {\bibinfo {volume} {75}},\
  \bibinfo {pages} {600} (\bibinfo {year} {1995})}\BibitemShut {NoStop}%
\bibitem [{\citenamefont {Polarski}\ and\ \citenamefont
  {Starobinsky}(1996)}]{polarskiSemiclassicalityDecoherenceCosmological1996}%
  \BibitemOpen
  \bibfield  {author} {\bibinfo {author} {\bibfnamefont {D.}~\bibnamefont
  {Polarski}}\ and\ \bibinfo {author} {\bibfnamefont {A.~A.}\ \bibnamefont
  {Starobinsky}},\ }\href {https://doi.org/10.1088/0264-9381/13/3/006}
  {\bibfield  {journal} {\bibinfo  {journal} {Classical and Quantum Gravity}\
  }\textbf {\bibinfo {volume} {13}},\ \bibinfo {pages} {377} (\bibinfo {year}
  {1996})}\BibitemShut {NoStop}%
\bibitem [{\citenamefont {Launay}\ \emph
  {et~al.}(2024{\natexlab{b}})\citenamefont {Launay}, \citenamefont
  {Rigopoulos},\ and\ \citenamefont {Shellard}}]{Launay:2024trh}%
  \BibitemOpen
  \bibfield  {author} {\bibinfo {author} {\bibfnamefont {Y.~L.}\ \bibnamefont
  {Launay}}, \bibinfo {author} {\bibfnamefont {G.~I.}\ \bibnamefont
  {Rigopoulos}},\ and\ \bibinfo {author} {\bibfnamefont {E.~P.~S.}\
  \bibnamefont {Shellard}},\ }\href@noop {} {\bibinfo {title} {{Quantitative
  classicality in cosmological interactions during inflation}}} (\bibinfo
  {year} {2024}{\natexlab{b}}),\ \Eprint {https://arxiv.org/abs/2412.16143}
  {arXiv:2412.16143 [gr-qc]} \BibitemShut {NoStop}%
\bibitem [{\citenamefont {Easther}\ and\ \citenamefont
  {Lim}(2006)}]{eastherStochasticGravitationalWave2006}%
  \BibitemOpen
  \bibfield  {author} {\bibinfo {author} {\bibfnamefont {R.}~\bibnamefont
  {Easther}}\ and\ \bibinfo {author} {\bibfnamefont {E.~A.}\ \bibnamefont
  {Lim}},\ }\href@noop {} {\bibfield  {journal} {\bibinfo  {journal} {Journal
  of Cosmology and Astroparticle Physics}\ }\textbf {\bibinfo {volume}
  {2006}}\bibinfo  {number} { (04)},\ \bibinfo {pages} {010}}\BibitemShut
  {NoStop}%
\bibitem [{\citenamefont {Aurrekoetxea}\ \emph
  {et~al.}(2023{\natexlab{a}})\citenamefont {Aurrekoetxea}, \citenamefont
  {Clough},\ and\ \citenamefont
  {Muia}}]{aurrekoetxeaOscillonFormationInflationary2023}%
  \BibitemOpen
\bibfield  {number} {  }\bibfield  {author} {\bibinfo {author} {\bibfnamefont
  {J.~C.}\ \bibnamefont {Aurrekoetxea}}, \bibinfo {author} {\bibfnamefont
  {K.}~\bibnamefont {Clough}},\ and\ \bibinfo {author} {\bibfnamefont
  {F.}~\bibnamefont {Muia}},\ }\href
  {https://doi.org/10.1103/PhysRevD.108.023501} {\bibfield  {journal} {\bibinfo
   {journal} {Physical Review D}\ }\textbf {\bibinfo {volume} {108}},\ \bibinfo
  {pages} {023501} (\bibinfo {year} {2023}{\natexlab{a}})},\ \Eprint
  {https://arxiv.org/abs/2304.01673} {arXiv:2304.01673 [astro-ph,
  physics:gr-qc, physics:hep-th]} \BibitemShut {NoStop}%
\bibitem [{\citenamefont {Deskins}\ \emph {et~al.}(2013)\citenamefont
  {Deskins}, \citenamefont {Jr},\ and\ \citenamefont
  {Caldwell}}]{deskinsGaugeFieldPreheating2013}%
  \BibitemOpen
  \bibfield  {author} {\bibinfo {author} {\bibfnamefont {J.~T.}\ \bibnamefont
  {Deskins}}, \bibinfo {author} {\bibfnamefont {J.~T.~G.}\ \bibnamefont {Jr}},\
  and\ \bibinfo {author} {\bibfnamefont {R.~R.}\ \bibnamefont {Caldwell}},\
  }\href {https://doi.org/10.48550/arXiv.1305.7226} {\bibinfo {title} {Gauge
  {{Field Preheating}} at the {{End}} of {{Inflation}}}} (\bibinfo {year}
  {2013}),\ \Eprint {https://arxiv.org/abs/1305.7226} {arXiv:1305.7226}
  \BibitemShut {NoStop}%
\bibitem [{\citenamefont {Battefeld}\ \emph {et~al.}(2009)\citenamefont
  {Battefeld}, \citenamefont {Battefeld},\ and\ \citenamefont
  {Giblin}}]{battefeldSuppressionParametricResonance2009}%
  \BibitemOpen
  \bibfield  {author} {\bibinfo {author} {\bibfnamefont {D.}~\bibnamefont
  {Battefeld}}, \bibinfo {author} {\bibfnamefont {T.}~\bibnamefont
  {Battefeld}},\ and\ \bibinfo {author} {\bibfnamefont {J.~T.}\ \bibnamefont
  {Giblin}},\ }\href {https://doi.org/10.1103/PhysRevD.79.123510} {\bibfield
  {journal} {\bibinfo  {journal} {Phys. Rev. D}\ }\textbf {\bibinfo {volume}
  {79}},\ \bibinfo {pages} {123510} (\bibinfo {year} {2009})}\BibitemShut
  {NoStop}%
\bibitem [{\citenamefont {Figueroa}\ \emph {et~al.}(2011)\citenamefont
  {Figueroa}, \citenamefont {{Garc{\'i}a-Bellido}},\ and\ \citenamefont
  {Rajantie}}]{figueroaTransversetracelessProjectionLattice2011}%
  \BibitemOpen
  \bibfield  {author} {\bibinfo {author} {\bibfnamefont {D.~G.}\ \bibnamefont
  {Figueroa}}, \bibinfo {author} {\bibfnamefont {J.}~\bibnamefont
  {{Garc{\'i}a-Bellido}}},\ and\ \bibinfo {author} {\bibfnamefont
  {A.}~\bibnamefont {Rajantie}},\ }\href
  {https://doi.org/10.1088/1475-7516/2011/11/015} {\bibfield  {journal}
  {\bibinfo  {journal} {Journal of Cosmology and Astroparticle Physics}\
  }\textbf {\bibinfo {volume} {2011}}\bibinfo  {number} { (11)},\ \bibinfo
  {pages} {015}}\BibitemShut {NoStop}%
\bibitem [{\citenamefont {Ota}\ \emph {et~al.}(2022)\citenamefont {Ota},
  \citenamefont {Macpherson},\ and\ \citenamefont
  {Coulton}}]{otaCovariantTransversetracelessProjection2022}%
  \BibitemOpen
\bibfield  {number} {  }\bibfield  {author} {\bibinfo {author} {\bibfnamefont
  {A.}~\bibnamefont {Ota}}, \bibinfo {author} {\bibfnamefont {H.~J.}\
  \bibnamefont {Macpherson}},\ and\ \bibinfo {author} {\bibfnamefont {W.~R.}\
  \bibnamefont {Coulton}},\ }\href
  {https://doi.org/10.1103/PhysRevD.106.063521} {\bibfield  {journal} {\bibinfo
   {journal} {Physical Review D}\ }\textbf {\bibinfo {volume} {106}},\ \bibinfo
  {pages} {063521} (\bibinfo {year} {2022})},\ \Eprint
  {https://arxiv.org/abs/2111.09163} {arXiv:2111.09163 [astro-ph,
  physics:gr-qc, physics:hep-th]} \BibitemShut {NoStop}%
\bibitem [{\citenamefont {Aurrekoetxea}\ \emph
  {et~al.}(2023{\natexlab{b}})\citenamefont {Aurrekoetxea}, \citenamefont
  {Clough},\ and\ \citenamefont {Lim}}]{aurrekoetxeaCTTKNewMethod2023}%
  \BibitemOpen
  \bibfield  {author} {\bibinfo {author} {\bibfnamefont {J.~C.}\ \bibnamefont
  {Aurrekoetxea}}, \bibinfo {author} {\bibfnamefont {K.}~\bibnamefont
  {Clough}},\ and\ \bibinfo {author} {\bibfnamefont {E.~A.}\ \bibnamefont
  {Lim}},\ }\href {https://doi.org/10.1088/1361-6382/acb883} {\bibfield
  {journal} {\bibinfo  {journal} {Classical and Quantum Gravity}\ }\textbf
  {\bibinfo {volume} {40}},\ \bibinfo {pages} {075003} (\bibinfo {year}
  {2023}{\natexlab{b}})},\ \Eprint {https://arxiv.org/abs/2207.03125}
  {arXiv:2207.03125 [gr-qc]} \BibitemShut {NoStop}%
\bibitem [{\citenamefont {Fergusson}\ \emph
  {et~al.}(2012{\natexlab{a}})\citenamefont {Fergusson}, \citenamefont
  {Liguori},\ and\ \citenamefont {Shellard}}]{fergussonCMBBispectrum2012}%
  \BibitemOpen
  \bibfield  {author} {\bibinfo {author} {\bibfnamefont {J.~R.}\ \bibnamefont
  {Fergusson}}, \bibinfo {author} {\bibfnamefont {M.}~\bibnamefont {Liguori}},\
  and\ \bibinfo {author} {\bibfnamefont {E.~P.~S.}\ \bibnamefont {Shellard}},\
  }\href {https://doi.org/10.1088/1475-7516/2012/12/032} {\bibfield  {journal}
  {\bibinfo  {journal} {Journal of Cosmology and Astroparticle Physics}\
  }\textbf {\bibinfo {volume} {2012}}\bibfield  {number} {\bibinfo  {number} {
  (12)},\ \bibinfo {pages} {032}},\ }\Eprint {https://arxiv.org/abs/1006.1642}
  {arXiv:1006.1642 [astro-ph]} \BibitemShut {NoStop}%
\bibitem [{\citenamefont {Fergusson}\ and\ \citenamefont
  {Shellard}(2007)}]{fergussonPrimordialNonGaussianityCMB2007}%
  \BibitemOpen
  \bibfield  {author} {\bibinfo {author} {\bibfnamefont {J.~R.}\ \bibnamefont
  {Fergusson}}\ and\ \bibinfo {author} {\bibfnamefont {E.~P.~S.}\ \bibnamefont
  {Shellard}},\ }\href {https://doi.org/10.1103/PhysRevD.76.083523} {\bibfield
  {journal} {\bibinfo  {journal} {Physical Review D}\ }\textbf {\bibinfo
  {volume} {76}},\ \bibinfo {pages} {083523} (\bibinfo {year}
  {2007})}\BibitemShut {NoStop}%
\bibitem [{\citenamefont {Akrami}\ \emph {et~al.}(2020)\citenamefont {Akrami}
  \emph {et~al.}}]{Planck:2018jri}%
  \BibitemOpen
  \bibfield  {author} {\bibinfo {author} {\bibfnamefont {Y.}~\bibnamefont
  {Akrami}} \emph {et~al.} (\bibinfo {collaboration} {Planck}),\ }\href
  {https://doi.org/10.1051/0004-6361/201833887} {\bibfield  {journal} {\bibinfo
   {journal} {Astron. Astrophys.}\ }\textbf {\bibinfo {volume} {641}},\
  \bibinfo {pages} {A10} (\bibinfo {year} {2020})},\ \Eprint
  {https://arxiv.org/abs/1807.06211} {arXiv:1807.06211 [astro-ph.CO]}
  \BibitemShut {NoStop}%
\bibitem [{\citenamefont {Fergusson}\ \emph
  {et~al.}(2012{\natexlab{b}})\citenamefont {Fergusson}, \citenamefont
  {Regan},\ and\ \citenamefont {Shellard}}]{Fergusson:2010ia}%
  \BibitemOpen
  \bibfield  {author} {\bibinfo {author} {\bibfnamefont {J.~R.}\ \bibnamefont
  {Fergusson}}, \bibinfo {author} {\bibfnamefont {D.~M.}\ \bibnamefont
  {Regan}},\ and\ \bibinfo {author} {\bibfnamefont {E.~P.~S.}\ \bibnamefont
  {Shellard}},\ }\href {https://doi.org/10.1103/PhysRevD.86.063511} {\bibfield
  {journal} {\bibinfo  {journal} {Phys. Rev. D}\ }\textbf {\bibinfo {volume}
  {86}},\ \bibinfo {pages} {063511} (\bibinfo {year} {2012}{\natexlab{b}})},\
  \Eprint {https://arxiv.org/abs/1008.1730} {arXiv:1008.1730 [astro-ph.CO]}
  \BibitemShut {NoStop}%
\bibitem [{\citenamefont {Hung}\ \emph {et~al.}(2019)\citenamefont {Hung},
  \citenamefont {Fergusson},\ and\ \citenamefont
  {Shellard}}]{hung2019advancingmatterbispectrumestimation}%
  \BibitemOpen
  \bibfield  {author} {\bibinfo {author} {\bibfnamefont {J.}~\bibnamefont
  {Hung}}, \bibinfo {author} {\bibfnamefont {J.~R.}\ \bibnamefont
  {Fergusson}},\ and\ \bibinfo {author} {\bibfnamefont {E.~P.~S.}\ \bibnamefont
  {Shellard}},\ }\href {https://arxiv.org/abs/1902.01830} {\bibinfo {title}
  {Advancing the matter bispectrum estimation of large-scale structure: a
  comparison of dark matter codes}} (\bibinfo {year} {2019}),\ \Eprint
  {https://arxiv.org/abs/1902.01830} {arXiv:1902.01830 [astro-ph.CO]}
  \BibitemShut {NoStop}%
\bibitem [{\citenamefont {Florio}\ and\ \citenamefont
  {Shellard}()}]{EF-EPS:Future}%
  \BibitemOpen
  \bibfield  {author} {\bibinfo {author} {\bibfnamefont {E.}~\bibnamefont
  {Florio}}\ and\ \bibinfo {author} {\bibfnamefont {E.~P.~S.}\ \bibnamefont
  {Shellard}},\ }\href@noop {} {\ }\Eprint {https://arxiv.org/abs/in
  preparation} {in preparation} \BibitemShut {NoStop}%
\bibitem [{\citenamefont {Frigo}\ and\ \citenamefont
  {Johnson}(2005)}]{frigoDesignImplementationFFTW32005}%
  \BibitemOpen
  \bibfield  {author} {\bibinfo {author} {\bibfnamefont {M.}~\bibnamefont
  {Frigo}}\ and\ \bibinfo {author} {\bibfnamefont {S.}~\bibnamefont
  {Johnson}},\ }\href {https://doi.org/10.1109/JPROC.2004.840301} {\bibfield
  {journal} {\bibinfo  {journal} {Proceedings of the IEEE}\ }\textbf {\bibinfo
  {volume} {93}},\ \bibinfo {pages} {216} (\bibinfo {year} {2005})}\BibitemShut
  {NoStop}%
\bibitem [{\citenamefont {Organization~(WMO)}\ \emph
  {et~al.}(1973)\citenamefont {Organization~(WMO)}, \citenamefont {Oliger},\
  and\ \citenamefont
  {Unions~(ICSU)}}]{organizationwmoMethodsApproximateSolution}%
  \BibitemOpen
  \bibfield  {author} {\bibinfo {author} {\bibfnamefont {W.~M.}\ \bibnamefont
  {Organization~(WMO)}}, \bibinfo {author} {\bibfnamefont {J.}~\bibnamefont
  {Oliger}},\ and\ \bibinfo {author} {\bibfnamefont {I.~C. o.~S.}\ \bibnamefont
  {Unions~(ICSU)}},\ }\href@noop {} {\bibinfo {title} {Methods for the
  {{Approximate Solution}} of {{Time Dependent Problems}}}} (\bibinfo {year}
  {1973})\BibitemShut {NoStop}%
\bibitem [{\citenamefont {Bona}\ \emph {et~al.}(2003)\citenamefont {Bona},
  \citenamefont {Ledvinka}, \citenamefont {Palenzuela},\ and\ \citenamefont
  {{\v Z}{\'a}{\v c}ek}}]{bonaGeneralcovariantEvolutionFormalism2003}%
  \BibitemOpen
  \bibfield  {author} {\bibinfo {author} {\bibfnamefont {C.}~\bibnamefont
  {Bona}}, \bibinfo {author} {\bibfnamefont {T.}~\bibnamefont {Ledvinka}},
  \bibinfo {author} {\bibfnamefont {C.}~\bibnamefont {Palenzuela}},\ and\
  \bibinfo {author} {\bibfnamefont {M.}~\bibnamefont {{\v Z}{\'a}{\v c}ek}},\
  }\href {https://doi.org/10.1103/PhysRevD.67.104005} {\bibfield  {journal}
  {\bibinfo  {journal} {Physical Review D}\ }\textbf {\bibinfo {volume} {67}},\
  \bibinfo {pages} {104005} (\bibinfo {year} {2003})}\BibitemShut {NoStop}%
\bibitem [{\citenamefont {Alic}\ \emph {et~al.}(2012)\citenamefont {Alic},
  \citenamefont {{Bona-Casas}}, \citenamefont {Bona}, \citenamefont
  {Rezzolla},\ and\ \citenamefont
  {Palenzuela}}]{alicConformalCovariantFormulation2012a}%
  \BibitemOpen
  \bibfield  {author} {\bibinfo {author} {\bibfnamefont {D.}~\bibnamefont
  {Alic}}, \bibinfo {author} {\bibfnamefont {C.}~\bibnamefont {{Bona-Casas}}},
  \bibinfo {author} {\bibfnamefont {C.}~\bibnamefont {Bona}}, \bibinfo {author}
  {\bibfnamefont {L.}~\bibnamefont {Rezzolla}},\ and\ \bibinfo {author}
  {\bibfnamefont {C.}~\bibnamefont {Palenzuela}},\ }\href
  {https://doi.org/10.1103/PhysRevD.85.064040} {\bibfield  {journal} {\bibinfo
  {journal} {Physical Review D}\ }\textbf {\bibinfo {volume} {85}},\ \bibinfo
  {pages} {064040} (\bibinfo {year} {2012})}\BibitemShut {NoStop}%
\end{thebibliography}%

\newpage
\appendix
\label{sec:appendix}
\section{Modulus/argument decomposition of the MS solutions}
\label{apdx:mod-arg-decomp}

In order to properly use the GRF generation method described in Sec.~\ref{subsec:ic-construction}, we must first break Eqns.~\eqref{eqn:MS-soln-field} and \eqref{eqn:MS-soln-velocity} into modulus-argument form:
\begin{eqnarray*}
    h_s(k) = M_h e^{i\theta_h},\\
    \dot{h}_s(k) = M_{\dot{h}} e^{i\theta_{\dot{h}}}.
\end{eqnarray*}
The moduli $M_h$ and $M_{\dot{h}}$ can be used as the basis for the Rayleigh draw. These moduli are given simply by the square root of the corresponding power spectrum. The argument can be found by splitting the complex exponential into sines and cosines using Euler's formula.
Let $k'=k/H_0.$ Then
\begin{eqnarray*}
    \theta_{h} &= \arctan\left(\frac{\cos(k') + k'\sin(k')}{k'\cos(k') - \sin(k')}\right)\\
    \\
    \theta_{\dot{h}} &= -\arctan\left(\frac{\cos(k')}{\sin(k')}\right).
\end{eqnarray*}
Note that one must use the arctan which takes into account the sign of the numerator and denominator.

\section{Construction of plus/cross basis in Fourier space}
\label{apdx:plus-cross-construction}

Since we will be generating the perturbation field $h_{ij}$ on the lattice, we will need a method of determining $\epsilon^{s}_{ij}$ at each point in the box, as a function of $\textbf{k}$, the three-dimensional wavenumber. 
For each point in the box, let $\textbf{k} = (i, j, k)$ represent the propagation of a particular gravitational wave, where $i, j,$ and $k$ are integers in the interval $(0, N-1)$.\footnote{In order to avoid confusion, I will denote the unit distance along the third axis in $\textbf{k}$ space as $k$, and write the magnitude of $\textbf{k}$ as $|\textbf{k}|$.} 
We now choose $\hat{\textbf{m}}$ and $\hat{\textbf{n}}$ such that they are orthogonal to $\hat{\textbf{k}}$ and to each other. Note that there exists a degeneracy in the choice of $\hat{\textbf{m}}$ and $\hat{\textbf{n}}$, and therefore we must choose a particular construction of these vectors.

Let $\hat{\textbf{m}} = \hat{\textbf{k}}\times(0, 0, k).$ Then $\hat{\textbf{m}}$ and $\hat{\textbf{n}}$ can be expressed in terms of the integer coordinates as 
\begin{eqnarray}
    \label{eqn:tensor-ics:region1}
    \nonumber \hat{\textbf{m}} &=& \frac{(j,-i,0)}{\sqrt{i^2+j^2}}\\
    \hat{\textbf{n}} &=& \frac{(i\ k, j\ k, i^2 + j^2)}{\sqrt{k^2(i^2+j^2) + (i^2 + j^2)^2}},
\end{eqnarray}
using the fact that $\hat{\textbf{n}} = \hat{\textbf{k}}\times\hat{\textbf{m}}.$ Note that this choice is only valid where $i, j, k > 0$, which I will call region 1. 
There are three other regions in which we will need to choose a different set of $\hat{\textbf{m}}$ and $\hat{\textbf{n}}:$ (2) $k > 0$ but $i = j = 0;$ (3) $k = 0$ but $j > 0$; and (4) $k = j = 0$ but $i > 0$. 
In two of these regions, the choice can be made trivially: for region 2, $\hat{\textbf{k}} = (0,0,1)$ and thus,
\begin{eqnarray}
    \label{eqn:tensor-ics:region2}
    \hat{\textbf{m}} &= (1,0,0),\ \hat{\textbf{n}} = (0,1,0).
\end{eqnarray}
In region 4, $\hat{\textbf{k}} = (1, 0, 0)$ and thus
\begin{eqnarray}
    \label{eqn:tensor-ics:region4}
    \hat{\textbf{m}} &= (0,1,0),\ \hat{\textbf{n}} = (0,0,1).
\end{eqnarray}
Now we simply choose a construction for region 3. In this case, let $\hat{\textbf{m}} = \hat{\textbf{k}}\times (i + j,0,0).$ Then
\begin{eqnarray}
    \label{eqn:tensor-ics:region3}
    \hat{\textbf{m}} &= (0,0,-1),\ \hat{\textbf{n}} = \frac{(-j, i, 0)}{\sqrt{i^2 + j^2}}.
\end{eqnarray}

Note that we can generate a new choice of decomposition by rotating $\hat{\textbf{m}}$ and $\hat{\textbf{n}}$. Let $\alpha\in (0,2\pi).$ 
Then the new basis 
\begin{eqnarray*}
    \hat{\textbf{m}}' &= \cos(\alpha)\hat{\textbf{m}} + \sin(\alpha)\hat{\textbf{n}},\ \ \ \hat{\textbf{n}}' = -\sin(\alpha)\hat{\textbf{m}} + \cos(\alpha)\hat{\textbf{n}}
\end{eqnarray*}
can also be used to construct a valid set of basis tensors $\epsilon^{+}_{ij}$ and $\epsilon^{\times}_{ij}$.

\section{Fourier conventions in FFTW}
\label{apdx:fourier-conventions}
The Fourier conventions given in Sec.~\ref{subsec:ics-formulation} represent the standard QFT normalisation of the Fourier transform. 
We implemented our Fourier transform with the FFTW package (Ref.~\cite{frigoDesignImplementationFFTW32005}), which is designed to transform ``unitless'' grids. 
This appendix lays out how we can transform to and from the FFTW conventions and the QFT conventions.

As we require that the perturbation field $h_{ij}(\textbf{x})$ be real, we use the ``r2c'' and ``c2r'' Fourier transforms available in the FFTW package. 
These functions neatly take care of the Hermitian symmetry necessary to form a real field.
As an example, the 1-dimensional ``c2r'' Fourier transform calculates 
\begin{equation}
    \label{eqn:fftw-ift}
    \mathbb{Y}_l = \sum_{j=0}^{N-1} \Delta j \ \mathbb{X}_j e^{2\pi (jl)\sqrt{-1}/N},
\end{equation}
where $j\in[0,N-1]$ and $l\in[0,N-1]$ are the unitless Fourier and configuration space axes, respectively.
We want to find a way to transform this into the standard discrete QFT transform,
\begin{equation}
    \label{eqn:qft-ift}
    \mathbb{Y}_{x_l} = \sum_{j=0}^{N-1} \Delta k_j \ \mathbb{X}_{k_j} \ e^{i k_j x_l}
\end{equation}
where $k_j = 2\pi j/L$ and $x_l = l\cdot D$ for $D = L/N.$ This will require two normalisations, one for the change of variables $j\rightarrow k$ in the IFT, and one for the change of variables $x\rightarrow l$ in the FT.

We start with the IFT, Eqn.~\eqref{eqn:fftw-ift}. We will change variables in the integral into the QFT unitful variables, then use the resulting norm as a multiplicative constant on the output of the FFTW algorithm. 

First, note that we can transform $j\rightarrow k$ using
\begin{equation*}
    k = \frac{2\pi j}{L} = \frac{2\pi j}{N\cdot D} = \mathbb{K}_j \frac{1}{D}
\end{equation*}
where we've defined the intermediate unitless mode magnitude to be $\mathbb{K}_j = 2\pi j/N.$ Thus
\begin{equation*}
     \mathbb{Y}(l) = \sum_{j=0}^{N-1} \Delta j \ \mathbb{X}_j e^{2\pi (jl)\sqrt{-1}/N} =  \frac{N}{2\pi} \sum_{j=0}^{N-1} \Delta \mathbb{K}_j \ \mathbb{X}_{\mathbb{K}_j} e^{i \mathbb{K}_j l}.
\end{equation*}
where the pre-factor comes from the discrete measure. Now we use
\begin{equation*}
    \Delta k_j = \frac{1}{D} \Delta \mathbb{K}_j \ \mbox{ and } \ l = \frac{x_l}{D}
\end{equation*}
to convert to a unitful mode integrand:
\begin{eqnarray*}
    \frac{N}{2\pi} \sum_{j=0}^{N-1} \Delta \mathbb{K}_j \ \mathbb{X}_{\mathbb{K}_j} e^{i \mathbb{K}_j l} &=& D \frac{N}{2\pi} \sum_{j=0}^{N-1} \Delta k_j \ \mathbb{X}_{k_j} e^{i \mathbb{K}_j x_l/D} \\
    &=& \frac{L}{2\pi} \sum_{j=0}^{N-1} \Delta k_j \ \mathbb{X}_{k_j} \ e^{i k_j x_l}.
\end{eqnarray*}
This gives
\begin{equation*}
    \mathbb{Y}_{x_l} \equiv \sum_{j=0}^{N-1} \Delta k_j \ \mathbb{X}_{k_j} \ e^{i k_j x_l} = \frac{2\pi}{L} \mathbb{Y}_l
\end{equation*}
where we've just moved the conversion factor to the LHS of the above set of equations.\footnote{Recall that $\mathbb{Y}(l)$ is the output of the ``c2r'' algorithm acting on a field given with a unitless argument.}

We can repeat this process for the ``r2c'' FT,
\begin{equation*}
    \mathbb{X}_j = \sum_{l=0}^{N-1} \Delta l \ \mathbb{Y}_l e^{-i(2\pi lj)/N}.
\end{equation*}
Using $\Delta l = \Delta x_l/D$ gives 
\begin{equation*}
    \mathbb{X}_j = \frac{1}{D} \sum_{l=0}^{N-1} \Delta x_l \ \mathbb{Y}_{x_l} e^{-i x_l (2\pi j/ND)} = \frac{1}{D} \sum_{l=0}^{N-1} \Delta x_l \ \mathbb{Y}_{x_l} e^{-i x_l k_j}
\end{equation*}
which means 
\begin{equation*}
    \mathbb{X}_{k_j} = \mathbb{X}_j \cdot D.
\end{equation*}

We can generalise this easily to three dimensions by raising the transformation factor to the power of the number of dimensions:
\begin{eqnarray}
    \label{eqn:fftw-qft-rescalings}
    \mathbb{Y}_{\textbf{x}_l} &=& \left(\frac{2\pi}{L}\right)^3 \mathbb{Y}_{\textbf{l}} \\
    \mathbb{X}_{\textbf{k}_j} &=& \left( \frac{L}{N} \right)^3 \cdot \mathbb{X}_{\textbf{j}}.
\end{eqnarray}

\section{The effect of KO dissipation on spectra}
\label{apdx:ko-dissipation}

NR simulations of highly curved spacetimes often use constraint damping to stay close to the desired solution surface. 
This can come in the form of Kreiss-Oliger dissipation (see Ref.~\cite{organizationwmoMethodsApproximateSolution}) or from the damping terms built into the CCZ4 formalism, another formulation of the ADM equations which is also used in GRChombo (see Refs.~\cite{bonaGeneralcovariantEvolutionFormalism2003,alicConformalCovariantFormulation2012a,radiaLessonsAdaptiveMesh2022} for reviews). 

However, dissipation can leave non-physical markers on perturbative systems, where precision in the solution is essential.
To demonstrate this effect, we ran the horizon-crossing case with various levels of KO dissipation, encapsulated by the unitless parameter $\sigma_{KO}$. We performed this run with a very light window, given by the window parameters 
\begin{equation*}
    k_* = 50\cdot \frac{2\pi}{L} \approx 0.1\ m_{pl}, \ \ \epsilon = 100,
\end{equation*}
where $N=64$, in order to best illustrate the effects of damping on high modes.

\begin{figure}[b]
    \centering
    \includegraphics[width=1.0\linewidth]{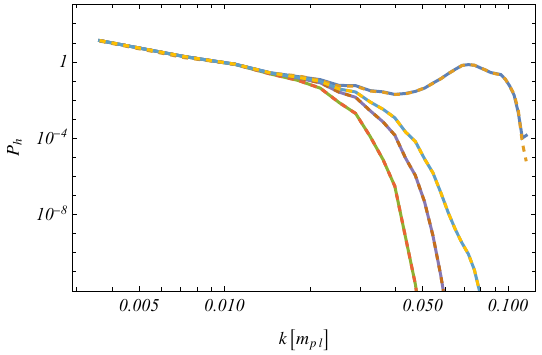}
    \caption{Comparison of the plus (solid) and cross (dashed) power spectra on the final slice, for $\sigma_{KO} = 0.3$ (green/red), $\sigma_{KO}=0.1$ (purple/brown), and $\sigma_{KO} = 0.05$ (light blue/yellow), compared to a run where KO dissipation was not used (dark blue/yellow).}
    \label{fig:KO-comp}
\end{figure}

Fig.~\ref{fig:KO-comp} shows the spectrum on the final slice, for parameter choices $\sigma_{KO} = 0.3$, $0.1$ and $0.05$.
We began by testing $\sigma_{KO}=0.3$, as this window value is typical of black hole simulations.
We note the distinct behaviour at high $k$, where a relatively high value of $\sigma_{KO}$ can strongly suppress some modes below the Nyquist bump, appearing here at $k \sim 0.08\ m_{pl}$.

Lower values of $\sigma_{KO}$ may be useful in damping power near the Nyquist frequency, as long as the parameter choice is appropriate for the physical spectrum evolved.
This parameter must be chosen carefully, however; KO dissipation is dynamical, and thus it's impact on the resulting signal would be more difficult to disentangle from the physical dynamical processes one has set out to study, compared to ``static'' damping techniques such as applying a window function on the initial data.
This consideration is often unnecessary for black hole simulations, as the modes of interest tend to be significantly lower than the range of KO dissipation.

\end{document}